\documentclass[twoside,11pt]{article}

\usepackage{blindtext}

\usepackage{jmlr2e}

\usepackage{amsmath}
\usepackage{xcolor}
\usepackage{bm}
\usepackage{amsfonts} 
\usepackage{bbm}
\usepackage{multirow}
\usepackage{hyperref}

\hypersetup{
    colorlinks=true, 
    linkcolor=blue,  
    citecolor=blue,   
    urlcolor=magenta 
}

\usepackage{lastpage}
\jmlrheading{23}{2024}{1-\pageref{LastPage}}{1/21; Revised 5/22}{9/22}{21-0000}{Chuyun Ye, Lixing Zhu, and Ruoqing Zhu}

\ShortHeadings{Consistent Order Determination of MDP}{Ye, Zhu, and Zhu}
\firstpageno{1}

\begin{document}

\title{Consistent Order Determination of Markov Decision Process}

\author{\name Chuyun Ye \email cyye@mail.bnu.edu.cn \\
       \addr School of Statistics\\
       Beijing Normal University\\
       Beijing, China
       \AND
       \name Lixing Zhu \email lzhu@bnu.edu.cn \\
       \addr Department of Statistics\\
       Beijing Normal University at Zhuhai\\
       Zhuhai, China
       \AND
       \name Ruoqing Zhu \email rqzhu@illinois.edu \\
       \addr Department of Statistics \\
       University of Illinois at Urbana-Champaign \\
       Champaign, USA
       }

\editor{My editor}

\maketitle

\begin{abstract}
The Markov assumption in Markov Decision Processes (MDPs) is fundamental in reinforcement learning, influencing both theoretical research and practical applications. Existing methods that rely on the Bellman equation benefit tremendously from this assumption for policy evaluation and inference. Testing the Markov assumption or selecting the appropriate order is important for further analysis. Existing tests primarily utilize sequential hypothesis testing methodology, increasing the tested order if the previously-tested one is rejected. However, This methodology cumulates type-I and type-II errors in sequential testing procedures that cause inconsistent order estimation, even with large sample sizes. To tackle this challenge, we develop a procedure that consistently distinguishes the true order from others. We first propose a novel estimator that equivalently represents any order Markov assumption. Based on this estimator, we thus construct a signal function and an associated signal statistic to achieve estimation consistency. Additionally, the curve pattern of the signal statistic facilitates easy visualization, assisting the order determination process in practice. Numerical studies validate the efficacy of our approach.
\end{abstract}


\section{Introduction}

Reinforcement learning (RL) involves an agent learning to make optimal decisions through interactions with an environment to achieve specific goals. It has seen successful applications in fields such as artificial intelligence, game theory, medicine, and finance. Central to many RL models is the Markov Decision Process (MDP) \citep{bellman1957markovian, puterman2014markov}, a mathematical framework that relies on the Markov property. This property asserts that the current state depends only on recent historical information rather than the entire history, which intuitively suggests that a policy based on recent information can be optimal compared to one based on the entire history \citep{puterman2014markov, shi2020does}. This characteristic simplifies both theoretical analysis and practical applications, making MDPs a powerful tool in RL.

Most of the existing works in RL assume that the joint process of state, action, reward sequences, which follows a first-order Markov property, indicating the current state depends only on the most recent state and action. This assumption greatly simplifies the Bellman equation \citep{sutton2018reinforcement}, facilitating the functional approximation of value or action-value functions in RL. However, in practice, many systems exhibit longer dependencies. For example, in medical settings, doctors make decisions based not only on a patient’s most recent treatment and health status but also on a sequence of prior information \citep{gottesman2019guidelines, komorowski2018artificial}. These higher-order Markov processes complicate RL modeling and solutions, necessitating more complex representations and computational methods to accurately capture extended temporal dependencies. Identifying the appropriate order is crucial for improving the robustness and performance of RL models in real-world applications, where first-order assumptions may not suffice.

Early methods for determining the order of Markov processes with continuous state spaces are closely related to lag selection in autoregressive models. In the context of nonparametric estimation of autoregressive models, researchers have developed model-free approaches to identify the true order or select the appropriate lag, using criteria such as Akaike Information Criterion, Bayesian Information Criterion, and Final Prediction Error. For more details, readers can refer to \cite{tong1975determination, auestad1990identification} and \cite{csiszar2000consistency}. Other approaches rely on hypothesis testing procedures. For instance, \cite{ait1997interest} and \cite{de2007testing} derived hypothesis tests based on the Chapman-Kolmogorov equation for first-order Markov processes, though these are not equivalent to directly testing the first-order Markov assumption. \cite{chen2012testing} proposed a test based on conditional characteristic functions to further explore the gap between the Chapman-Kolmogorov equation and the first-order Markov assumption. These methods can be extended to testing higher-order processes by applying suitable transformations to the observed data.

To determine the order, a straightforward approach is to perform a sequential testing procedure on a series of null hypotheses until a null hypothesis is not rejected. By carefully choosing the significance level, one can also determine the order of the Markov process. \cite{shi2020does} applied this idea by consecutively conducting tests until the $k$-th null hypothesis is not rejected. However, two issues remain unresolved in their methodology. First, the method is based on a function of the order that is zero under the null hypothesis, but under the alternative hypothesis, the function cannot be ensured to be non-zero. This property does not provide an equivalency to the $k$-order Markov assumption. As a result, each test in the sequential testing procedure may not be sensitive to all alternative hypotheses, meaning that in some cases, the order cannot be accurately determined. In other words, each test is not omnibus. Second, it is well known that the estimated order determined by the sequential testing procedure cannot converge to the true order in probability. This is because each test has a nonzero type-I error, leading to a non-zero cumulative type-I error in the sequential testing procedure. See \cite{zhou2023testing} for a relevant reference.

To address these issues, we propose a novel method to consistently estimate the order of a Markov process. Instead of performing sequential tests of increasing orders, our estimator is based on a signal statistic that achieves consistent selection. The proposed estimation procedure comprises two steps: detecting violations of the $k$-order Markov assumption and distinguishing orders smaller than the true order, $k_0$.

In the first step, we define a function that equivalently represents the $k$-order Markov assumption, which means that it is zero when the $k$-order Markov assumption holds and nonzero when it is violated. In the second step, we construct a signal function and the corresponding signal statistic to consistently estimate $k_0$. The curve of the signal function exhibits a pattern where the true order is the largest minimizer among all zeros, with all subsequent values equal to the constant 1 beyond $k_0$. This feature facilitates accurate identification of the true order and can also assist in visually identifying the order in practice. The empirical signal statistic maintains this pattern asymptotically, ensuring the method's consistency. The implementation involves searching for the minimizer in a sequence, making the algorithm efficient and fast. We also derive the consistency of the estimation. To the best of our knowledge, this is the first attempt to obtain a consistent estimator for this problem.

The remainder of this paper is organized as follows. Section \ref{preliminary} formally introduces the order determination problem. Section \ref{signal} outlines the estimation procedure and discusses the consistency of the estimation. Section \ref{simulation} presents numerical studies to validate our method. Finally, Section \ref{RealData} demonstrates the application of our approach using a real dataset.

\section{Distinguishing the Markov Assumption} \label{preliminary}

\subsection{The Markov Assumption}

Suppose we have a trajectory of length $T$, $\{ (S_t, A_t, R_t) \}_{t=1}^T$, observed from a discrete-time Markov decision process $\mathcal{M}_0$. For any $t \in \{1, \ldots, T\}$, $(S_t, A_t, R_t)$ is a state-action-reward triplet, where $S_t$ is a $p$-dimensional state variable, and $A_t$ is a scalar action variable. For notational simplicity, let $X_t := \left( S_t^\top, A_t \right)^\top$, and let $(X_m)_{t_1}^{t_2} := \left(X_{t_1}^\top,\cdots,X_{t_2}^\top\right)^\top$ denote the state-action information within the epoch interval $[t_1, t_2]$. The action $A_t$ is generated from a stationary policy $\pi\left(\cdot \,\vert\, (S_m)_{t-k_0}^{t}\right)$, which depends solely on the historical information $(S_m)_{t-k_0}^{t}$. The reward $R_t$ comes from a deterministic function of $(S_t, A_t, S_{t+1})$. Assume that we collect $N$ independent and identically distributed (i.i.d.) observations from $\mathcal{M}_0$ with length $T$, i.e., $\{ \mathcal{D}_j \}_{j=1}^N = \left\{ \{ (S_{j,t}, A_{j,t}, R_{j,t}) \}_{t=1}^T \right\}_{j=1}^N$.

$\mathcal{M}_0$ is a Markov decision process if and only if the probability distribution of $S_{t+1}$ given $(X_m)_{1}^{t}$ depends only on part of the historical information. More specifically, $\mathcal{M}_0$ is a $k_0$-order Markov decision process if and only if the following assumption holds.

\begin{itemize}
	\item [(A1)] ($k_0$-order Markov Assumption) For any $t>k_0$,
	\begin{equation*}\label{conditional independence}
		S_{t+1} \perp (X_m)_{1}^{t-k_0} \, \vert \,  (X_m)_{t-k_0+1}^{t},
	\end{equation*}
	or equivalently,
	\begin{equation}\label{Markov assumption 1}
		\mathrm{Pr} \Big( S_{t+1} \in \mathcal{S} \, \big\vert \, (X_m)_{1}^{t} \Big) = \mathrm{Pr} \Big( S_{t+1} \in \mathcal{S} \, \big\vert \,  (X_m)_{t-k_0+1}^{t} \Big),
	\end{equation}
	almost surely, for arbitrary Borel set $\mathcal{S} \in \mathbb{S}$, where $\mathbb{S}$ is the invariant event space of $S_t$ for any $t$.
\end{itemize}

We further assume that the conditional probability measure is time-invariant. 
\begin{itemize}
        \item [(A2)] For any $t\geq k_0$, 
        \begin{equation}
            \mathrm{Pr} \Big( S_{t+1} \in \mathcal{S} \, \big\vert \, (X_m )_{t-k_0+1}^{t} = \tilde{x} \Big) = \mathcal{P} \big( \mathcal{S} \, \big\vert \, \tilde{x} \big),
        \end{equation}
        where $\mathcal{P} \big( \cdot \, \big\vert \, \tilde{x} \big)$ is a time-invariant probability measure for any given $\tilde{x}$.
\end{itemize}

The following two assumptions are also commonly made for Markov chains \citep{mengersen1996rates}:

\begin{itemize}
    \item [(A3)] The state sequence $\{S_t\}$ is strictly stationary.
    \item [(A4)] The sequence $\{X_t\}$ is geometrically ergodic.
\end{itemize}

Since $A_t$ is generated from a distribution determined by ${ (S_m) }_{t-k_0}^t$, Assumption (A3) further implies that the sequence ${ X_t }$ is also strictly stationary, a common assumption in existing reinforcement learning research. Moreover, based on Theorem 3.7 of \cite{bradley2005basic}, Assumptions (A3) and (A4) together imply that ${ X_t }$ is exponentially $\beta$-mixing, which limits the dependency of the sequence over time. Geometric ergodicity or exponential $\beta$-mixing is also assumed by \cite{antos2008learning}, \cite{dai2018sbeed}, \cite{wang2023projected}, \cite{shi2024statistically}, among others. This assumption on mixing is weaker than other commonly used assumptions, including i.i.d. observed data \citep{laroche2019safe, dai2020coindice} and uniform mixing \citep{bhandari2018finite, zou2019finite, luckett2020estimating}. It is important to emphasize that both (A3) and (A4) are widely used assumptions regarding stationarity and time dependence and are satisfied in numerous models. For example, consider a vector autoregression model of order $k_0$ (VAR($k_0$)) for the state variable $S_t$ (readers can refer to \cite{lutkepohl2005new} for details). With mild assumptions for the error terms and the action variable $A_t$, the corresponding process is a $k_0$-order MDP that satisfies (A3) and (A4). 

\subsection{An Equivalent Condition and Its Related Function}

In this subsection, we first introduce an equivalent condition for the $k_0$-order Markov assumption and then propose a new criterion to measure the deviation of a Markov process from a particular order Markov assumption. Based on this proposed criterion, we define a statistic and derive its asymptotic properties. This statistic serves as a foundation for constructing a signal statistic used for order determination.

To identify the true order of a Markov decision process $\mathcal{M}_0$, we need to measure and detect deviations of $\mathcal{M}_0$ from non-$k_0$-order Markov assumptions. A straightforward approach would be to verify Equation (\ref{Markov assumption 1}) for different candidate values of order $k$. However, verifying an equation involving an infinite set of historical information is practically infeasible when $T$ approaches infinity. Therefore, we consider an equivalent condition for the $k_0$-order Markov assumption, as presented in the following theorem.

\begin{theorem} \label{EquivalentTheorem}
	The $k_0$-order Markov assumption with Equation (\ref{Markov assumption 1}) is equivalent to
\begin{align}
	& \mathrm{E} \Big[ \exp \left( i \mu^\top S_{t+q+k_0} + i \nu^\top X_{t-1} \right) \, \big\vert \, ( X_m )_{t}^{t+q+k_0-1} \Big] \nonumber \\
	= & \mathrm{E} \Big[ \exp \left( i \mu^\top S_{t+q+k_0} \right) \, \big\vert \,  ( X_m )_{t+q}^{t+q+k_0-1} \Big] \cdot \mathrm{E} \Big[ \exp \left( i \nu^\top X_{t-1} \right) \, \big\vert \,  ( X_m )_{t}^{t+q+k_0-1} \Big] \label{Equi}
\end{align}
almost surely, for any $q \geq 0$, $2\leq t \leq T-q-k_0$, $\mu \in \mathbb{R}^p$, and $\nu \in \mathbb{R}^{p+1}$.
\end{theorem}

When $k_0=1$, this theorem is analogous to Theorem 1 of \cite{shi2020does}. Through this theorem, we can measure the deviation of $\mathcal{M}_0$ from any $k$-order Markov assumption by quantifying the extent of the violation,
\begin{align}
	\bigg\vert & \mathrm{E} \Big[ \exp \left( i \mu^\top S_{t+q+k} + i \nu^\top X_{t-1} \right) \, \big\vert \,  ( X_m )_{t}^{t+q+k-1} \Big] \nonumber \\
	& - \mathrm{E} \Big[ \exp \left( i \mu^\top S_{t+q+k} \right) \, \big\vert \,  ( X_m )_{t+q}^{t+q+k-1} \Big] \cdot \mathrm{E} \Big[ \exp \left( i \nu^\top X_{t-1} \right) \, \big\vert \,  ( X_m )_{t}^{t+q+k-1} \Big] \bigg\vert^2, \label{Diff}
\end{align}
for arbitrary choice of $q$, $\mu$, and $\nu$. For any $k < k_0$, there exists some $q$, $\mu$, and $\nu$, such that the squared module (\ref{Diff}) is strictly greater than 0. Otherwise, for $k \geq k_0$, the squared module is equal to 0 for any $q$, $\mu$, and $\nu$, as well as the one taking expectation with respect to $(X_m)_{t}^{t+q+k-1}$:
\begin{align}
    \Gamma^{(k,q)}(\mu,\nu) = & \mathrm{E} \Bigg\{ \bigg\vert \mathrm{E} \Big[ \exp \left( i \mu^\top S_{t+q+k} + i \nu^\top X_{t-1} \right) \Big\vert ( X_m )_{t}^{t+q+k-1} \Big] \nonumber \\
    & \quad - \mathrm{E} \Big[ \exp \left( i \mu^\top S_{t+q+k} \right) \Big\vert  ( X_m )_{t+q}^{t+q+k-1} \Big]  \mathrm{E} \Big[ \exp \left( i \nu^\top X_{t-1} \right) \Big\vert  ( X_m )_{t}^{t+q+k-1} \Big] \bigg\vert^2 \Bigg\}. \label{ModOfSub}
\end{align}

The following assumption helps define the types of violations that can be detected, with further explanations provided afterward:

\begin{itemize}
    \item [(A5)] For any $k < k_0$, there exists $q$, such that $\Gamma^{(k,q)}(\mu,\nu)$ is strictly larger than 0 in a non-zero measure set for $\mu$ and $\nu$.
\end{itemize}

Under this assumption, we can distinguish between violations of Equation (\ref{Equi}) that occur only on a zero-measure set and those that occur on a set with a non-zero measure. The former type of violation is typically not of interest, making Assumption (A5) reasonable. The violations constrained by Assumption (A5) are detectable by $\Gamma^{(k,q)}(\mu,\nu)$. It is also important to note that in the subsequent section on estimation, this assumption can be slightly relaxed. With Assumption (A5) in place, it follows that there exists a $q$, $\mu$, and $\nu$ such that $\Gamma^{(k,q)}(\mu,\nu) > 0$ for any $k < k_0$. The performance difference of $\Gamma^{(k,q)}(\mu,\nu)$ between $k < k_0$ and $k \geq k_0$ is crucial for distinguishing these two ranges of $k$, making $\Gamma^{(k,q)}(\mu,\nu)$ a fundamental quantity in our analysis.

\begin{remark} We also highlight an important difference between the way we measure violations and the approach taken by \cite{shi2020does}. Briefly, they defined the following function to measure deviations from the $k$-order Markov assumption:
\begin{equation*}
    \bigg\vert \mathrm{E}\Big\{ \left[ \exp\left(i \mu^\top S_{t+q+k}\right) - \mathrm{E}\left[\exp\left(i \mu^\top S_{t+q+k}\right)\vert (X_{m})_{t+q}^{t+q+k-1}\right] \right] \exp\left(i \nu^\top X_{t-1}\right) \Big\} \bigg\vert_{\infty}.
\end{equation*}

Here, $\vert c \vert_{\infty} := \max{ \vert \text{Re}(c) \vert, \vert \text{Im}(c)\vert }$ for any complex number $c$. Both this function and $\Gamma^{(k,q)}(\mu,\nu)$ are derived from Theorem \ref{EquivalentTheorem}. The main difference between these measuring ways lies in whether or not the marginal expectation is taken to eliminate certain conditional expectation structures. Since conditional characteristic functions take values in $[-1,1]\times [-1,1]$, our function does not take the marginal expectation. This ensures that $\Gamma^{(k,q)}(\mu,\nu) = 0$ for any $q$, $\mu$, and $\nu$ is equivalent to the $k$-order Markov assumption under Assumption (A5), whereas the function defined in \cite{shi2020does} does not guarantee this equivalence, and thus is not omnibus against all alternative hypotheses \citep{zhou2023testing}.
\end{remark}

When it comes to constructing the empirical form of $\Gamma^{(k,q)}$, we turn the problem into a regression task to estimate the conditional expectation structures in $\Gamma^{(k,q)}$. To maintain the independence between estimated functions and arguments, we randomly separate the accessible sample set $\{ 1,\cdots,N \}$ into two sets,  $\mathcal{L}$ and $\mathcal{L}^C:=\{ 1,\cdots, N \} - \mathcal{L}$. When $N$ is an even integer, the sizes of two sub-samples are $N/2$, otherwise, they are $(N+1)/2$ and $(N-1)/2$. But for notation simplicity, we assume an even $N$ in the remaining parts of this paper. Let $g(Z;\mu,\nu)$ denote any conditional characteristic functions in Equation (\ref{ModOfSub}), where $Z$ represents $(X_m)_{t}^{t+q+k-1}$ or $(X_m)_{t+q}^{t+q+k-1}$. The estimator of $g$ constructed by $\{\mathcal{D}_j\}_{j \notin \mathcal{L}}$ is denoted by $\hat{g}^{(-\mathcal{L})}$. As several methods are available in the literature to define consistent estimators, we do not discuss a specific method to define estimators for conditional characteristic functions but make the following restriction on the convergence rate of estimators.

\begin{itemize}
	\item [(A6)] For any $\mu\in\mathbb{R}^p$ and $\nu\in\mathbb{R}^{p+1}$,
	\begin{equation*}
		\mathrm{E} \left\{ \Big[ g(Z;\mu,\nu) - \hat{g}^{(-\mathcal{L})}(Z;\mu,\nu) \Big]^2 \, \Big\vert \, \{ \mathcal{D}_j\}_{j \not\in \mathcal{L}} \right\}=o_p\left( \frac{1}{\log(NT)} \right),
	\end{equation*}
	where the conditional expectation is taken with respect to $Z$ following the distribution of $(X_m)_{1}^{q+k}$ or $(X_m)_{1}^{k}$.
\end{itemize}

In the numerical studies, we use random forests \citep{breiman2001random} to estimate all conditional characteristic functions here. According to \cite{goehry2020random}, (A6) can be satisfied under some mild conditions, when the involving sequence is exponentially $\beta$-mixing, which can be directly derived from (A3) and (A4).

Let $\mathrm{E}_n^{(-\mathcal{L})}$ represent the estimator of corresponding conditional characteristic function constructed by $\{\mathcal{D}_j\}_{j \notin \mathcal{L}}$. We adopt the empirical form of $\Gamma^{(k,q)}$ as
\begin{align*}
    & \Gamma^{(k,q)}_{N,T}(\mu,\nu) \\
    = & \frac{1}{N_\mathcal{L}(T-q-k+1)}  \\
    &  \times \sum_{j\in\mathcal{L}} \sum_{t=1}^{T-q-k+1} \Bigg\{ \bigg\vert \mathrm{E}_n^{(-\mathcal{L})} \Big[ \exp \left( i \mu^\top S_{j,t+q+k} + i \nu^\top X_{j,t-1} \right) \Big\vert ( X_{j,m} )_{t}^{t+q+k-1} \Big]  \\
    & \quad - \mathrm{E}_n^{(-\mathcal{L})} \Big[ \exp \left( i \mu^\top S_{j,t+q+k} \right) \Big\vert  ( X_{j,m} )_{t+q}^{t+q+k-1} \Big]  \mathrm{E}_n^{(-\mathcal{L})} \Big[ \exp \left( i \nu^\top X_{j,t-1} \right) \Big\vert  ( X_{j,m} )_{t}^{t+q+k-1} \Big] \bigg\vert^2 \Bigg\}.
\end{align*}
where $N_{\mathcal{L}} = \# \mathcal{L}$. To alleviate the impact from  arbitrarily chosen values $q$, $\mu$, and $\nu$, we consider $(Q+1)$ choices of $q$ ($q=0,\cdots,Q$), and $B$ randomized directions for $\mu$ and $\nu$ respectively ($\{ (\mu_b, \nu_b) \}_{b=1}^B$). With some pre-set positive integers $B$ and $Q$, we propose the following statistic to capture the violation of $\Gamma^{(k,q)}_{N,T}$ from $0$:
\begin{equation*}
	\Pi^{(k)}_{N,T} = \max_{0 \le q \le Q} \max_{1 \le b \le B}\Gamma^{(k,q)}_{N,T}(\mu_b,\nu_b),
\end{equation*}
where $\mu_b$ and $\nu_b$ are i.i.d. $p$-dimensional and $(p+1)$-dimensional standard Normal random variables, respectively, for $b=1,\cdots,B$. We further define
\begin{equation*}
	R^{(k)} = \max_{0 \leq q \leq Q} \sup_{\mu \in \mathbb{R}^p, \nu \in \mathbb{R}^{p+1}} \mathrm{E}\left[ \Gamma^{(k,q)}(\mu,\nu) \right].
\end{equation*}
 Theorem 1 and Assumption (A5) lead to: when $k_0>1$
\begin{equation}\label{R}
	 \begin{cases}
		R^{(k)} > 0, \qquad & 1 \leq k < k_0, \\
		R^{(k)} = 0, \qquad & k_0 \le  k \le K.
	\end{cases}
\end{equation}
 When $k_0=1$, $R^{(k)}=0$ for $k_0=1 \le k\le K$. In the following theorem, we state the convergence result from $\Pi^{(k)}_{N,T}$ to $R^{(k)}$. 

\begin{theorem} \label{Stage1Thm}
	Suppose that $B=\big( (NT)^{c^\star} \big)$ for any finite $c^\star>0$ and $Q+K \leq \max(\rho_0 T, T-2)$ for some constant $\rho_0<1$, and Assumptions (A1)-(A6) hold.  Then
	\begin{enumerate}
		\item uniformly over $k_0 \leq k \leq K$,
		\begin{equation*}
			\max_{0\leq q \leq Q} \max_{1 \leq b \leq B} \sqrt{N_\mathcal{L}(T-q-k+1)}  \Gamma_{N,T}^{(k,q)}(\mu_b, \nu_b) = o_p(1),
		\end{equation*}
		 as $N \to \infty$ or $T \to \infty$;
		\item uniformly over $1 \leq k \leq K$,
		 \begin{equation*}
			\left\vert \Pi^{(k)}_{N,T} - R^{(k)} \right\vert = O_p\left( \sqrt{\frac{\ln(NT)}{NT}} \right),
		\end{equation*}
		 as $N \to \infty$ or $T \to \infty$ where $R^{(k)}$ is defined in (\ref{R}).
	\end{enumerate}
\end{theorem}
The second result in Theorem \ref{Stage1Thm} plays a crucial role in order determination later as it distinguishes the values of $\Pi^{(k)}_{N,T}$ before and after the true order $k_0$. Based on this fact, we propose a signal statistic based on this fact in the next section.

\section{Signal Function and Statistic for Order Determination} \label{signal}

In this section, we propose a novel estimation procedure for the true order $k_0$. To illustrate the estimation nature of this procedure, we begin by introducing a signal function and the corresponding empirical form, a signal statistic.

We start with the convergence nature of $\Pi_{N,T}^{(k)}$ for any $1 \leq k \leq K$. According to Theorem \ref{Stage1Thm}, $\Pi_{N,T}^{(k)}$ converges to $R^{(k)}$, which inspires further investigation of $R^{(k)}$. Recall that the sequence ${ R^{(k)} }_{1 \leq k \leq K}$ has the property that $R^{(k)} > 0$ only when $1 \leq k \leq k_0 - 1$. One intuitive idea to identify $k_0$ is to look for the smallest $k$ such that $R^{(k)} = 0$. However, since the magnitude of $R^{(k)}$ varies under different models when $k = 1, \dots, k_0 - 1$, it may be difficult to distinguish between $R^{(k)}$ and 0 when $k = 1, \dots, k_0 - 1$.

To cope with this scalar-variate problem, we can further consider the ratio sequence as $\{ R^{(k)} / R^{(k-1)} \}_{k=1}^K$. To maintain the good definition, let $R^{(0)}=1$. Therefore, the ratio sequence satisfies that
\begin{equation*}
    \frac{R^{(k)}}{R^{(k-1)}} \begin{cases}
        > 0, \qquad  & 1 \leq k \leq k_0-1; \\
        = 0, \qquad  & k = k_0; \\
        = 0/0, \qquad  & k_0+1 \leq k \leq K.
    \end{cases}
\end{equation*}
Here we adopt the notation ‘$0/0$’ to represent the fraction with both the numerator and denominator equal to 0. So far, we can identify $k_0$ by utilizing $R^{(k_0)}/R^{(k_0-1)}=0$. The remaining problem lies on that the quantity $0/0$ is undefined. To this end, we introduce a positive ridge value $c_{N,T}$. Adding the ridge value to both numerator and denominator simultaneously can effectively avoid the existence of $0/0$ but directly turn it into $(0+c_{N,T})/(0+c_{N,T}) = 1$. The ridge value $c_{N,T}$ is required not only positive but also sufficiently small, such that it would not have great impact on the value of $(R^{(k)}+c_{N,T})/(R^{(k-1)}+c_{N,T})$ when $1 \leq k \leq k_0$. To maintain the consistency of the estimator of $k_0$, $c_{N,T}$ will be demanded to decay as $N$ or $T$ goes to infinity, shortly. Consider a signal function defined on $1 \le k \le K$ as
\begin{equation*}
    \Omega^{(N,T)}(k) = \frac{(R^{(k)})^\eta+c_{N,T}}{(R^{(k-1)})^\eta+c_{N,T}}, 
\end{equation*}
where $R^{(0)}=1$. For $k_0>1$, 
 \begin{equation*}\label{R1}
	\Omega^{(N,T)}(k) = \frac{(R^{(k)})^\eta+ c_{N,T}}{(R^{(k-1)})^\eta+ c_{N,T}} = \begin{cases}
		\frac{(R^{(k)})^\eta+ c_{N,T}}{(R^{(k-1)})^\eta+ c_{N,T}}, \qquad & 1 \leq k < k_0, \\
        \frac{0 + c_{N,T}}{(R^{(k-1)})^\eta+ c_{N,T}}, \qquad & k=k_0,\\
		\frac{c_{N,T}}{c_{N,T}}, \qquad & k_0 <  k \le K.
	\end{cases}
\end{equation*}
And as $N$ or $T$ goes to infinity,
\begin{equation*}
    \Omega^{(N,T)}(k) \to C(k) \begin{cases}
        >0, \qquad & 1 \leq k < k_0, \\
        =0, \qquad & k=k_0,\\
        =1, \qquad & k_0 <  k \le K.
    \end{cases}
\end{equation*}
 For $k_0=1$, 
  \begin{equation*}
	\Omega^{(N,T)}(k) \to C(k) \begin{cases}
            = 0, \qquad & k=1,\\
		= 1, \qquad & 1 <  k \le K.
	\end{cases}
\end{equation*}
as $N$ or $T$ goes to infinity. Notice that apart from $c_{N,T}$, we also replace $R^{(k)}/R^{(k-1)}$ with the $\eta$-powered one, $\left( R^{(k)}/R^{(k-1)} \right)^\eta$. We usually choose $\eta>1$ to amplify the magnitude difference between the $\Omega^{(N,T)}(k)$ at $k>k_0$ and at other locations $k$s. On the other hand, in practical performing, larger value of $\eta$ may also tend to overestimate $k_0$, as the magnitude difference of $\Omega^{(N,T)}(k)$ among $k>k_0$. For practical use, we recommend $\eta=3$.

Therefore, the true order $k_0$ can be identified by utilizing that $\Omega^{(N,T)}(k_0)$ converges to 0.

Based on this signal function, we adopt its  empirical form to define the signal statistic $\Omega_{N,T}^{(k)}$. Remind that Theorem~\ref{Stage1Thm} yields that $\Pi^{(k)}_{N,T}$ are consistent estimators of $R^{(k)}$ uniformly over $1 \le k \le K$. We define $\Pi^{(0)}_{N,T}=1$. For $1 \leq k \leq K$, we define the signal statistic $\Omega^{(k)}_{N,T}$ as
\begin{equation}
	\Omega^{(k)}_{N,T} := \frac{(\Pi^{(k)}_{N,T})^\eta + c_{N,T}}{(\Pi^{(k-1)}_{N,T})^\eta + c_{N,T}}.
\end{equation}

However, one should notice that directly adding a constant ridge $c_{N,T}$ may lead to a ratio without scale-invariant property. Furthermore, in some small-scale numerical experiments, $\Pi^{(k)}_{N,T}$ often takes very small values, because it is a function of the difference between two conditional characteristic functions, which makes it hard to select $c_{N,T}$. Therefore, we further consider a semi-data-driven ridge as 
\begin{equation*}
    \tilde c_{N,T}=c_{N,T}\left(\max_{1\le k\le K} \Pi^{(k)}_{N,T}\right)^\eta
\end{equation*}
Equivalently, we define the final signal statistic as
\begin{equation*}
\tilde{\Omega}^{(k)}_{N,T} = \frac{\left(\Pi^{(k)}_{N,T}\right)^\eta + \tilde c_{N,T}}{\left(\Pi^{(k-1)}_{N,T}\right)^\eta + \tilde c_{N,T}}= \frac{\left(\Pi^{(k)}_{N,T}/\max_{1\le k\le K} \Pi^{(k)}_{N,T}\right)^\eta +  c_{N,T}}{\left(\Pi^{(k-1)}_{N,T}/\max_{1\le k\le K} \Pi^{(k)}_{N,T}\right)^\eta +  c_{N,T}}.
\end{equation*}
Therefore, one can regard the signal statistic with ridge $\tilde{c}_{N,T}$ as the signal statistic with re-scaled $\Pi^{(k)}_{N,T}$ and the original ridge $c_{N,T}$. Based on the signal statistic sequence $\{\tilde{\Omega}^{(k)}_{N,T}\}_{k=1}^K$, and the property of the signal function above-mentioned, we propose the estimator of $k_0$ as
\begin{equation}
	\hat{k}_\tau = \arg \max_k \Big\{ k: \tilde\Omega^{(k)}_{N,T} \le \tau \Big\}.
\end{equation}
Based of the consistency of $\Omega^{(k)}_{N,T}$, we can further derive the consistency of $\hat{k}_\tau$. We formulate this result in the following theorem.


\begin{theorem}\label{consistency}
	Suppose that the ridge value $c_{N,T}$ satisfies that $c_{N,T}\to 0$, $c_{N,T} \sqrt{NT}/\sqrt{\ln(NT)} \to \infty$, and $c_{N,T}/(R^{(k)}/\max_{1\le k\le K} R^{(k)})^3 \to 0$, as $N$ or $T$ goes to infinity. Under all the conditions in Theorem~\ref{Stage1Thm}, 
	\begin{equation*}
		\lim_{N \to \infty \text{ or } T \to \infty} \mathrm{Pr} \left( \hat{k}_\tau=k_0 \right) = 1.
	\end{equation*}
\end{theorem}

\begin{remark} The consistency of $\hat{k}_\tau$ in this theorem relies on the selection of two tuning parameters.  As mentioned before,  the ridge $c_{N,T}$ shall be small enough not to dominate over the non-zero $(R^{(k_0-1)}/\max_{1\le k\le K} R^{(k)})^3$. As $c_{N,T}$ goes to zero, in this sense, $(R^{(k_0-1)}/\max_{1\le k\le K} R^{(k)})^3$ can also go to zero which can be regarded as local models. For $\tau$, any fixed value between 0 and 1 can be used in theory. But practically, due to the random oscillation of the signal statistic, a large $\tau$ might cause determining a large $\hat k$, whereas a small $\tau$ might result in a small $\hat k$. As a compromise, we recommend $\tau=0.5$. Note the selection of $\tau$ is not data-driven, which deserves further study.
\end{remark}

\section{Numerical Study}\label{simulation}

To further test the performance of our proposed method, we conduct numerical experiments to evaluate the proposed estimation method in finite sample settings. It is worthwhile to mention that the validity of the proposed method remains invariant to the inherent true order $k_0$. To consider similar true order settings as in the real data example in the next section, we set all models with the true order $k_0 = 2$. Specifically, we conduct numerical studies under the following two models.

\noindent\textbf{Model 1:}
\begin{equation}
	S_{t} =  \frac{4}{5} \big[A_{t-1} \tilde{\bm{I}}_p + (1-A_{t-1}) {\bm{I}}_p\big] S_{t-2} + \varepsilon_t,
\end{equation}
\textbf{Model 2:}
\begin{equation}
	S_{t} =  \sqrt{\frac{[\ln(NT)]^{3}}{{NT}}}
 \big[A_{t-1} \tilde{\bm{I}}_p + (1-A_{t-1}) {\bm{I}}_p\big] S_{t-2} + \frac{2}{5} \big[A_{t-1} \tilde{\bm{I}}_p + (1-A_{t-1}) {\bm{I}}_p\big] S_{t-1} + \varepsilon_t.
\end{equation}
Here, the $p$-dimensional error term $\varepsilon_t$ is generated from $\mathcal{N}_p\big( 0, (3/p){\bm{I}}_p \big)$ independently. Let $A_t$ be determined by $S_t$ through a deterministic function $a(\cdot)$ as 
\begin{equation*} a(s) = \mathbbm{1} \Big( \bm{1}_p^\top s > 0 \Big), \end{equation*} 
where $\bm{1}_p$ is a $p$-dimensional unit vector. $\bm{I}_p$ and $\tilde{\bm{I}}_p$ denote a $p \times p$ identity matrix and a $p \times p$ anti-diagonal matrix with anti-diagonal elements equal to 1, respectively. We consider two cases of dimension $p$, specifically $p=3$ and $p=6$. Both Models 1 and 2 are stationary and ergodic Markov decision processes with true order $k_0 = 2$. Specifically, Model 2 can be regarded as a local 2-order MDP, which converges to a 1-order MDP as $NT$ goes to infinity. Numerical studies under Model 2 aim to examine the validity of our method in local models.

We perform two methods, our proposed estimation procedure and the sequential-hypothesis-testing-style method proposed by \cite{shi2020does}, to estimate the true order $k_0$. For our proposed method, we set $K=6$, $Q=5$, $B=\lfloor(NT)^{1/4}\rfloor$, $\tau=0.5$, and $c_{N,T}=0.1[\ln(N_\mathcal{L}T)]^{(a/2+1)}(N_\mathcal{L}T)^{-a/2}$ as the pre-set parameters. For the method proposed by \cite{shi2020does}, we conduct $K$ hypothesis tests with $B$ sets of independent $\mu$ and $\nu$. This setting applies in both this numerical study section and the real data analysis section. Moreover, the classic hypothesis testing method proposed by \cite{chen2012testing} should have been included in this comparison. However, the nonparametric estimation utilized by this method is no longer valid when working with our settings ($p=3$ and $p=6$). Furthermore, a small-scale numerical experiment with this method indicates that the computation is much more heavily loaded than those in the aforementioned two methods. For these reasons, we decided not to further perform simulations with the method proposed by \cite{chen2012testing}.

To examine consistency as $NT$ approaches infinity, we consider three pairs of $(N, T)$: $(6, 450)$, $(12, 450)$, and $(18, 450)$. We adopt these settings to mimic the long-sequence and small-sample-size nature of mobile health datasets, such as the OhioT1DM dataset. We repeat each experiment 500 times. In Tables \ref{SimulationResult_m1p3} and \ref{SimulationResult_m2p3}, we display the Monte Carlo means, mean square errors (MSEs), and empirical distributions of $\hat{k}_\tau-k_0$.

\begin{table}[h!]
\caption{Monte Carlo Means, MSEs, and Empirical Distribution of $\hat{k}$ in Model 1}
\label{SimulationResult_m1p3}
\centering
\begin{tabular}{cccccccccc}
\hline
\multirow{2}{*}{Method}     & \multirow{2}{*}{$(N,T)$} & \multirow{2}{*}{Mean} & \multirow{2}{*}{MSE} & \multicolumn{6}{c}{$\hat{k}-k_0$}      \\ \cline{5-10} & & & & -1 & 0 & 1 & 2 & 3 & 4 \\ \hline
\multicolumn{10}{c}{$p=3$}
\\ \hline
\multirow{3}{*}{Proposed} & (6,450) & 2.012 & 0.072 & 0.00 & 0.98 & 0.00 & 0.02 & 0.00 & 0.00 \\
 & (12,450) & 2.000 & 0.000 & 0.00 & 1.00 & 0.00 & 0.00 & 0.00 & 0.00 \\
 & (18,450) & 2.000 & 0.000 & 0.00 & 1.00 & 0.00 & 0.00 & 0.00 & 0.00 \\ \cline{2-10} 
\multirow{3}{*}{Shi (0.01)} & (6,450) & 2.036 & 0.108 & 0.00 & 0.99 & 0.00 & 0.01 & 0.00 & 0.00 \\
 & (12,450) & 2.020 & 0.056 & 0.00 & 0.99 & 0.00 & 0.01 & 0.00 & 0.00 \\
 & (18,450) & 2.032 & 0.076 & 0.00 & 0.99 & 0.00 & 0.01 & 0.00 & 0.00 \\ \hline
\multicolumn{10}{c}{$p=6$}
\\ \hline
\multirow{3}{*}{Proposed} & (6,450) & 2.008 & 0.016 & 0.00 & 1.00 & 0.00 & 0.00 & 0.00 & 0.00 \\
 & (12,450) & 2.036 & 0.072 & 0.00 & 0.98 & 0.00 & 0.02 & 0.00 & 0.00 \\
 & (18,450) & 2.004 & 0.008 & 0.00 & 1.00 & 0.00 & 0.00 & 0.00 & 0.00 \\ 
 \cline{2-10} 
\multirow{3}{*}{Shi (0.01)} & (6,450) & 1.112 & 0.952 & 0.91 & 0.08 & 0.00 & 0.01 & 0.00 & 0.00 \\
 & (12,450) & 4.020 & 4.115 & 0.00 & 0.00 & 0.00 & 0.99 & 0.01 & 0.01 \\
 & (18,450) & 4.012 & 4.072 & 0.00 & 0.00 & 0.00 & 0.99 & 0.00 & 0.01 \\ 
 \hline
\end{tabular}
\end{table}

\begin{table}[h!]
\caption{Monte Carlo Means, MSEs, and Empirical Distribution of $\hat{k}$ in Model 2 
}
\label{SimulationResult_m2p3}
\centering
\begin{tabular}{cccccccccc}
\hline
\multirow{2}{*}{Method}     & \multirow{2}{*}{$(N,T)$} & \multirow{2}{*}{Mean} & \multirow{2}{*}{MSE} & \multicolumn{6}{c}{$\hat{k}-k_0$} \\
\cline{5-10} 
 & & & & -1 & 0 & 1 & 2 & 3 & 4 \\
\hline
\multicolumn{10}{c}{$p=3$} \\
\hline
\multirow{3}{*}{Proposed} & (6,450) & 1.898 & 0.410 & 0.08 & 0.91 & 0.00 & 0.01 & 0.00 & 0.00 \\
 & (12,450) & 2.004 & 0.008 & 0.00 & 1.00 & 0.00 & 0.00 & 0.00 & 0.00 \\
 & (18,450) & 2.000 & 0.000 & 0.00 & 1.00 & 0.00 & 0.00 & 0.00 & 0.00 \\ 
 \cline{2-10} 
\multirow{3}{*}{Shi (0.01)} & (6,450) & 2.076 & 0.221 & 0.01 & 0.95 & 0.02 & 0.01 & 0.01 & 0.00 \\
 & (12,450) & 2.022 & 0.050 & 0.00 & 0.99 & 0.01 & 0.00 & 0.00 & 0.00 \\
 & (18,450) & 2.042 & 0.102 & 0.00 & 0.98 & 0.01 & 0.01 & 0.00 & 0.00 \\ 
\hline
\multicolumn{10}{c}{$p=6$} \\
\hline
\multirow{3}{*}{Proposed} & (6,450) & 1.988 & 0.100 & 0.00 & 0.99 & 0.00 & 0.01 & 0.00 & 0.00 \\
 & (12,450) & 2.000 & 0.000 & 0.00 & 1.00 & 0.00 & 0.00 & 0.00 & 0.00 \\
 & (18,450) & 2.164 & 0.172 & 0.00 & 0.84 & 0.16 & 0.00 & 0.00 & 0.00 \\ 
\cline{2-10} 
\multirow{3}{*}{Shi (0.01)} & (6,450) & 1.244 & 0.916 & 0.82 & 0.13 & 0.04 & 0.01 & 0.00 & 0.00 \\
 & (12,450) & 3.299 & 1.961 & 0.00 & 0.00 & 0.72 & 0.26 & 0.01 & 0.01 \\
 & (18,450) & 2.462 & 0.542 & 0.00 & 0.57 & 0.40 & 0.02 & 0.00 & 0.00 \\ 
\hline
\end{tabular}
\end{table}

For our proposed method, in both Models~1 and 2, the Monte Carlo means of $\hat{k}_\tau$ are close to $k_0=2$ and the Monte Carlo MSEs are close to 0, with growing $N,T$. This phenomenon helps verify the consistency of $\hat{k}_\tau$. For empirical distributions of $\hat{k}_\tau-k_0$, the frequency of $\hat{k}_\tau-k_0=0$ approaches 1, as $NT$ approaches infinity.   Further, for  Model 2 with weak signals being regarded as a local model, our method still works well,  the proportion of $|\hat k_{\tau}-k_0|\le 1$ is greater than $0.98$, and thus it verifies $\hat{k}_\tau$'s consistency.  We shall also mention that the results we displayed are not sensitive to the choice of $c_{N,T}$, with the condition in Theorem \ref{consistency} satisfied. 

For the estimation proposed by \cite{shi2020does}, notice that the estimation performance would be different by choosing different significant levels in the sequential testing procedure. Following the recommendation of \cite{shi2020does}, we have conducted the procedure with significant level as 0.01, 0.05, and 0.1. And we select the significant level with best performance to be displayed in this section, which is 0.01. In settings with $p=3$, our proposed method performs slightly better than Shi's method. When $p=6$, Shi's method tends to overestimate the true order with larger sample sizes, in both global and local models. For example, in Model 1 with $p=6$, Shi's method is more likely to accept the estimator $\hat{k}=4$. The cause of this phenomenon may be the accumulation of type-I and type-II error through the sequential testings. From this aspect, our proposed method has been shown consistent in finite sample cases, which is consistent with our theoretical analysis.

\begin{figure}[h!]
\centering
\includegraphics[width=0.8 \textwidth]{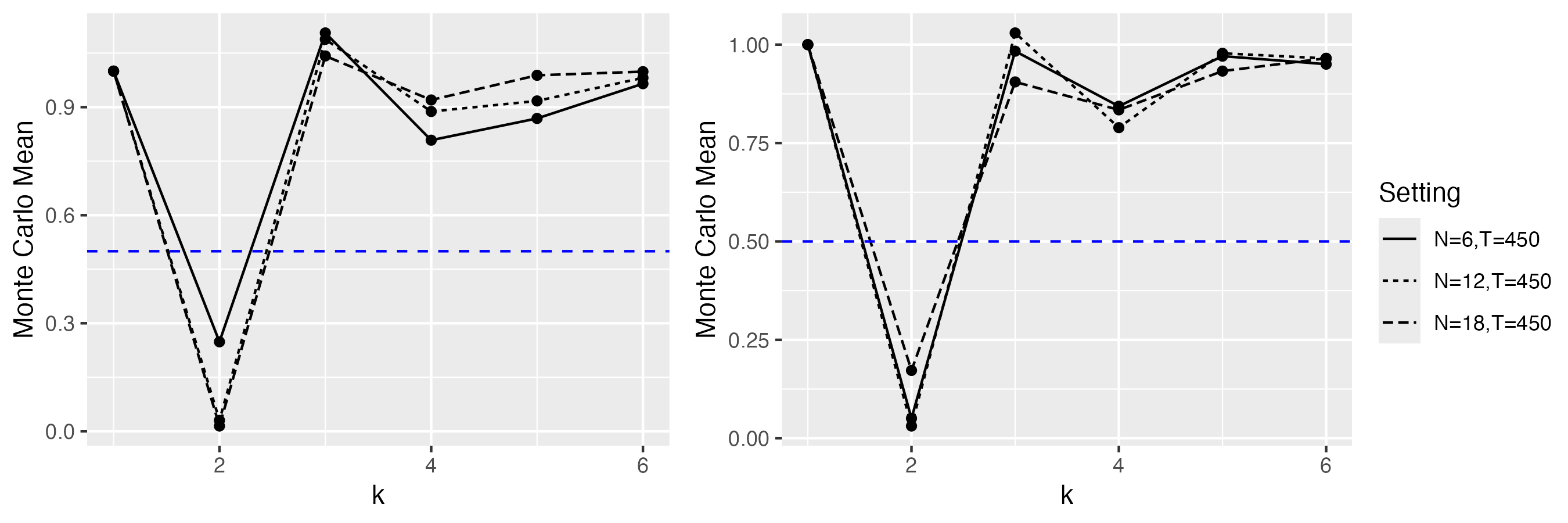}
\caption{Reported Values of $\Omega_{N,T}^{(k)}$ and $\tau$ in Model~1 (left: $p=3$; right: $p=6$)}
\label{Omega_p3p6mean}
\end{figure}

Furthermore, we display the Monte Carlo means of $\Omega_{N,T}^{(k)}$ in Figure \ref{Omega_p3p6mean} of Model 1 to visualize the curve of the sequence $\{\Omega_{N,T}^{(k)}\}_{k=1}^K$. As shown in the figures, the Monte Carlo means of $\Omega_{N,T}^{(k)}$ tend to approach 0 when $k=k_0=2$, and return to values near 1 when $k>k_0$. We can also capture the convergence of $\Omega_{N,T}^{(k)}$ with growing $NT$ with any fixed $k$. As an empirical example, Figure \ref{Omega_p3p6mean} helps demonstrate the identification nature of our proposed method. 

\section{Real Data Application}\label{RealData}

In this section, we apply the proposed method to the real dataset to show the capacity of the proposed method in real data analysis. We adopt the OhioT1DM dataset (\cite{marling2020ohiot1dm}) as an example. This dataset includes continuous glucose monitoring (CGM) records of 12 individuals with type 1 diabetes, their insulin injection doses, and other baseline covariates recorded sequentially. The OhioT1DM dataset is an informative dataset for analysing the relationship between the glucose level and daily activities of patients with type 1 diabetes, as well as predicting the glucose level, and finding out the optimal insulin injection regime for each patient. For each patient, the observed data in this dataset can be regarded as a decision process, with the insulin dose as the decision variable, and other covariates as state variables. While modeling, assuming this decision process as a Markov decision process would be of great interest, since this assumption can greatly reduce the complexity of statistic models. This leads to the problem of estimating the true order of this Markov decision process.   

In our analysis, for each timestamp $t$, we consider a 3-dimensional state variable $S_t$, including average continuous glucose monitoring data per hour, carbohydrate intake per hour, and exercise intensity per hour. We define the insulin infusion $A_t$ as the discretized amount of insulin injection dose per hour. In this dataset, after pre-processing, the sample size and length of the process shall be $N=12$ and $T=925$.

\begin{figure}[h!]
\centering
\includegraphics[width=0.4 \textwidth]{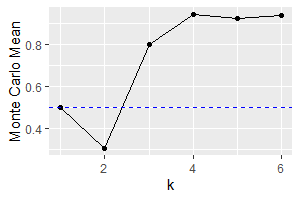}
\caption{Reported Values of $\Omega_{N,T}^{(k)}$ and $\tau$}
\label{RealData_plot}
\end{figure}

We report the values of $\Omega_{N,T}^{(k)}$ in Table \ref{RealData_table}. The curve of the sequence $\{\Omega_{N,T}^{(k)}\}_{k=1}^K$ is visualized in Figure \ref{RealData_plot}.
\begin{table}[h!]
\centering
\caption{Values of $\Omega_{N,T}^{(k)}$}
\label{RealData_table}
\begin{tabular}{|c|c|c|c|c|c|}
\hline
$k=1$   & $k=2$   & $k=3$   & $k=4$   & $k=5$   & $k=6$ \\ \hline
0.500 & 0.302 & 0.798 & 0.944 & 0.925 & 0.935  \\ \hline
\end{tabular}
\end{table}

From both the table and plot, we can directly derive that $\hat{k}_{0.5}=2$. The estimation indicates that the decision process for each patient in this experiment is a Markov decision process with order 2. Based on this estimation, the result supports statistical models including historical information for the last 2 hours, which hopefully, can improve the accuracy of further analysis for this dataset. 

To further demonstrate the capacity of our method with datasets like the OhioT1DM dataset, we estimate a vector autoregressive model for the real data, with lag $k_0=2$. Notice that this model is more complex than the models we considered in Section \ref{simulation}. The estimated model is reported in the supplementary material. With this model, we regenerate data with the same $(N,T)$ settings as Section \ref{simulation}, to mimic the data structure of the OhioT1DM dataset. The experiment is conducted for 500 repetitions. The corresponding results are reported in Table \ref{RealData_hatk_performance}. The reported result shows that with $NT$ approaching infinity, the Monte Carlo mean of $\hat{k}_{0.5}$ tends to be 2, while the Monte Carlo MSE approaches 0. This result further verifies the consistency of $\hat{k}_{0.5}$ in finite sample cases. Moreover, the phenomenon that the MSE of $\hat{k}_{0.5}$ is equal to 0 in cases ($N=12, T=450$) and ($N=18, T=450$), suggests the reliability of our method when analysing the OhioT1DM dataset. 

\begin{table}[h!]
\centering
\caption{Empirical distribution of the estimated order $\hat k_{\tau}$}
\label{RealData_hatk_performance}
\begin{tabular}{ccccccccc}
\hline
\multirow{2}{*}{$(N,T)$} & \multirow{2}{*}{Mean} & \multirow{2}{*}{MSE} & \multicolumn{6}{c}{$\hat{k}_\tau-k_0$} \\ 
\cline{4-9} 
 & & & -1 & 0 & 1 & 2 & 3 & 4 \\ \hline
(6,450) & 2.274& 0.274& 0.000& 0.726& 0.274&  0.000& 0.000& 0.000\\
(12,450) & 2.000& 0.000& 0.000& 1.000& 0.000& 0.000& 0.000& 0.000\\
(18,450) & 2.000& 0.000& 0.000& 1.000& 0.000& 0.000& 0.000& 0.000\\ \hline
\end{tabular}
\end{table}

\section{Discussion}

In this paper, we propose a novel estimation procedure for the true order of Markov decision processes. Based on the curve nature of the signal statistic, our proposed estimator is easy to visualize. Therefore, we can obtain a more accurate estimation. The consistency of the proposed estimator has been proven in this work. By numerical studies and real data analysis, the estimator has been verified to be valid in finite sample cases.

For tuning parameters $c_{N,T}$, we provide a semi-data-driven selection method, in order to avoid the scale-variant problem caused by directly adding $c_{N,T}$. For $\tau$, small $\tau$ may lead to underestimation, while large $\tau$ may lead to overestimation. In this work, we recommend $\tau=0.5$ based on the rule of thumb. However, it is more interpretable to determine $\tau$ by a data-driven way. This will be one of the directions of our future research.


\acks{
This research was supported by NSFC grants (NSFC12131006, NSFC12471276) and an NSF grant (\#2210657). The names of coauthors are in alphabetical order with equal contributions.}

\appendix
\section{Technical Proof}
\subsection{Proof of Theorem \ref{EquivalentTheorem}}

Recall that the $k_0$-order Markov assumption  in Equation \eqref{Markov assumption 1} is
\begin{equation}
	\mathrm{Pr} \Big( S_{t+1} \in \mathcal{S} \, \big\vert \, \{ X_m \}_{m \le t} \Big) = \mathrm{Pr} \Big( S_{t+1} \in \mathcal{S} \, \big\vert \, \{ X_m \}_{m=t-k_0+1}^{t} \Big),
\end{equation}
for any integer $t$.

We can immediately derive that Equation (\ref{Markov assumption 1}) implies that
\begin{align}
	& \mathrm{E} \Big[ \exp \left( i \mu^\top S_{t+q+k_0} + i \nu^\top X_{t-1} \right) \, \big\vert \, ( X_m )_{t}^{t+q+k_0-1} \Big] \nonumber \\
	= & \mathrm{E} \Big[ \exp \left( i \mu^\top S_{t+q+k_0} \right) \, \big\vert \,  ( X_m )_{t+q}^{t+q+k_0-1} \Big] \cdot \mathrm{E} \Big[ \exp \left( i \nu^\top X_{t-1} \right) \, \big\vert \,  ( X_m )_{t}^{t+q+k_0-1} \Big], \label{Equi_Appx}
\end{align}
for any $t \in \mathbb{Z}$, $q \in \mathbb{N}$, $\mu \in \mathbb{R}^p$, and $\nu \in \mathbb{R}^{p+1}$.

We then prove the reverse side of the equivalence. Suppose that Equation (\ref{Equi_Appx}) holds for any $t$, $q$, $\mu$ and $\nu$. First let $q=0$, then we have
\begin{align}
	& \mathrm{E} \Big[ \exp \left( i \mu^\top S_{t+1} + i \nu^\top X_{t-k_0} \right) \, \big\vert \, ( X_m )_{t-k_0+1}^{t} \Big] \nonumber \\
	= & \mathrm{E} \Big[ \exp \left( i \mu^\top S_{t+1} \right) \, \big\vert \, ( X_m )_{t-k_0+1}^{t} \Big] \cdot \mathrm{E} \Big[ \exp \left( i \nu^\top X_{t-k_0} \right) \, \big\vert \,  ( X_m )_{t-k_0+1}^{t} \Big] \label{q=0},
\end{align}
for any $t$, $\mu$, and $\nu$. Equation (\ref{q=0}) indicates that
\begin{equation}
	S_{t+1} \perp X_{t-k_0} \, \vert \, (X_m)_{t-k_0+1}^{t}
\end{equation}
for any $t$. Then let $q=1$. We have
\begin{align}
	& \mathrm{E} \Big[ \exp \left( i \mu^\top S_{t+1} + i \nu^\top X_{t-k_0-1} \right) \, \big\vert \, ( X_m )_{t-k_0}^{t} \Big] \nonumber \\
	= & \mathrm{E} \Big[ \exp \left( i \mu^\top S_{t+1} \right) \, \big\vert \,  ( X_m )_{t-k_0+1}^{t} \Big] \cdot \mathrm{E} \Big[ \exp \left( i \nu^\top X_{t-k_0-1} \right) \, \big\vert \,  ( X_m )_{t-k_0}^{t} \Big]. \label{q=1}
\end{align}
Multiplying $\exp(i v^\top X_{t-k_0})$ on both sides of Equation (\ref{q=1}), we can then obtain
\begin{align}
	& \mathrm{E} \Big[ \exp \left( i \mu^\top S_{t+1} + i \nu^\top X_{t-k_0-1} + i v^\top X_{t-k_0} \right) \, \big\vert \, ( X_m )_{t-k_0}^{t} \Big] \nonumber \\
	= & \mathrm{E} \Big[ \exp \left( i \mu^\top S_{t+1} \right) \, \big\vert \,  ( X_m )_{t-k_0+1}^{t} \Big] \cdot \mathrm{E} \Big[ \exp \left( i \nu^\top X_{t-k_0-1} + i v^\top X_{t-k_0} \right) \, \big\vert \,  ( X_m )_{t-k_0}^{t} \Big]. \label{q=1_mul}
\end{align}
Then taking the expectation of both sides of Equation (\ref{q=1_mul}) conditioning on $( X_m )_{t+1}^{t+k_0}$, we can obtain that
\begin{align*}
	& \mathrm{E} \Big[ \exp \left( i \mu^\top S_{t+1} + i \nu^\top X_{t-k_0-1} + i v^\top X_{t-k_0} \right) \, \big\vert \, ( X_m )_{t-k_0+1}^{t} \Big] \\
	= & \mathrm{E} \Big[ \exp \left( i \mu^\top S_{t+1} \right) \, \big\vert \,  ( X_m )_{t-k_0+1}^{t} \Big] \cdot \mathrm{E} \Big[ \exp \left( i \nu^\top X_{t-k_0-1} + i v^\top X_{t-k_0} \right) \, \big\vert \,  ( X_m )_{t-k_0+1}^{t} \Big].
\end{align*}
which implies that 
\begin{equation}
	S_{t+1} \perp (X_{t-k_0-1},X_{t-k_0}) \, \vert \, ( X_m )_{t-k_0+1}^{t}
\end{equation}
for any $t$. Similarly we can obtain that
\begin{equation*} 
    S_{t+1} \perp (X_{t-k_0-q+1}, \cdots, X_{t-k_0}) \, \vert \, ( X_m )_{t-k_0+1}^{t},
\end{equation*}
for any $q$ and $t$. Following the induction idea, we can derive that 
\begin{equation*}
    S_{t+1} \perp \{ X_j \}_{j \le t-k_0} \, \vert \, ( X_m )_{t-k_0+1}^{t},
\end{equation*}
and Equation (\ref{Markov assumption 1}). \hfill\BlackBox

\subsection{Proof of Theorem \ref{Stage1Thm}}

Define the following functions as
\begin{align}
	& g_{1,R} \big( \tilde{x}, \mu, \nu; k, q \big) = \mathrm{E} \Big[ \cos \big( \mu^\top S_{t+q+k} + \nu^\top X_{t-1} \big) \, \big\vert \, ( X_m )_{t}^{t+q+k-1} = \tilde{x} \Big],\nonumber \\
	& g_{1,I} \big( \tilde{x}, \mu, \nu; k, q \big) = \mathrm{E} \Big[ \sin \big( \mu^\top S_{t+q+k} + \nu^\top X_{t-1} \big) \, \big\vert \, ( X_m )_{t}^{t+q+k-1} = \tilde{x} \Big],\nonumber \\
	& g_{2,R} \big( \tilde{x}, \mu; k, q \big) = \mathrm{E} \Big[ \cos \big( \mu^\top S_{t+q+k} \big) \, \big\vert \, ( X_m )_{t+q}^{t+q+k-1} = \tilde{x} \Big], \nonumber \\
	& g_{2,I} \big( \tilde{x}, \mu; k, q \big) = \mathrm{E} \Big[ \sin \big( \mu^\top S_{t+q+k} \big) \, \big\vert \,( X_m )_{t+q}^{t+q+k-1} = \tilde{x} \Big], \nonumber \\
	& g_{3,R} \big( \tilde{x}, \nu; k, q \big) = \mathrm{E} \Big[ \cos \big( \nu^\top X_{t-1} \big) \, \big\vert \, ( X_m )_{t}^{t+q+k-1} = \tilde{x} \Big], \nonumber \\
	& g_{3,I} \big( \tilde{x}, \nu; k, q \big) = \mathrm{E} \Big[ \sin \big( \nu^\top X_{t-1} \big) \, \big\vert \, ( X_m )_{t}^{t+q+k-1} = \tilde{x} \Big], \label{gfuncs}
\end{align}
such that 
\begin{align*}
	& \mathrm{E} \Big[ \exp \left( i \mu^\top S_{t+q+k} + i \nu^\top X_{t-1} \right) \, \big\vert \,  ( X_m )_{t}^{t+q+k-1} \Big] \\
	= & g_{1,R}\left( ( X_m )_{t}^{t+q+k-1},\mu,\nu;k,q \right) + i \cdot g_{1,I}\left( ( X_m )_{t}^{t+q+k-1},\mu,\nu;k,q \right), \\
	& \mathrm{E} \Big[ \exp \left( i \mu^\top S_{t+q+k} \right) \, \big\vert \,  ( X_m )_{t+q}^{t+q+k-1} \Big] \\
	= & g_{2,R} \left( ( X_m )_{t+q}^{t+q+k-1},\mu;k,q \right) + i \cdot g_{2,I} \left( ( X_m )_{t+q}^{t+q+k-1},\mu;k,q \right), \\
	& \mathrm{E} \Big[ \exp \left( i \nu^\top X_{t-1} \right) \, \big\vert \,  ( X_m )_{t}^{t+q+k-1} \Big] \\
	= & g_{3,R} \left( ( X_m )_{t}^{t+q+k-1},\nu;k,q \right) + i \cdot g_{3,I} \left( ( X_m )_{t}^{t+q+k-1},\nu;k,q \right).
\end{align*}
We also notice that in these newly defined conditional expectations, $\tilde{x}$'s are vectors with different dimensions. But for notation simplicity without confusion, we use the same notation $\tilde{x}$. Furthermore, Subtraction (\ref{Diff}) can be rewritten as
\begin{align}
	& \Big[g_{1,R} \big((X_m)_{t}^{t+q+k-1},\mu,\nu;k,q\big) - g_{2,R} \big((X_m)_{t+q}^{t+q+k-1},\mu,\nu;k,q\big) g_{3,R} \big((X_m)_{t}^{t+q+k-1},\mu,\nu;k,q\big) \nonumber \\  
	& \quad + g_{2,I} \big((X_m)_{t+q}^{t+q+k-1},\mu,\nu;k,q\big) g_{3,I} \big((X_m)_{t}^{t+q+k-1},\mu,\nu;k,q\big) \Big]^2 \nonumber \\
	& \, + \Big[g_{1,I} \big((X_m)_{t}^{t+q+k-1},\mu,\nu;k,q\big) - g_{2,I} \big((X_m)_{t+q}^{t+q+k-1},\mu,\nu;k,q\big) g_{3,R} \big((X_m)_{t}^{t+q+k-1},\mu,\nu;k,q\big) \nonumber \\ 
	& \, \qquad  - g_{2,R} \big((X_m)_{t+q}^{t+q+k-1},\mu,\nu;k,q\big) g_{3,I} \big((X_m)_{t}^{t+q+k-1},\mu,\nu;k,q\big) \Big]^2. \label{ModOfSub_RI}
\end{align}
According to Theorem \ref{EquivalentTheorem}, we can derive that, for any $k \ge k_0$, expression (\ref{ModOfSub_RI}) should be equal to 0, for arbitrary $t$, $q$, $\mu$, and $\nu$, as well as its marginal expectation with respect to $(X_m)_{t}^{t+q+k-1}$:
\begin{align*}
	\Gamma^{(k,q)}(\mu,\nu) = & \mathrm{E} \bigg\{ \Big[g_{1,R} \big((X_m)_{t}^{t+q+k-1},\mu,\nu;k,q\big) \\
	& \qquad \quad - g_{2,R} \big((X_m)_{t+q}^{t+q+k-1},\mu,\nu;k,q\big) g_{3,R} \big((X_m)_{t}^{t+q+k-1},\mu,\nu;k,q\big)  \\  
	& \qquad \quad + g_{2,I} \big((X_m)_{t+q}^{t+q+k-1},\mu,\nu;k,q\big) g_{3,I} \big((X_m)_{t}^{t+q+k-1},\mu,\nu;k,q\big) \Big]^2  \\
	& \qquad + \Big[g_{1,I} \big((X_m)_{t}^{t+q+k-1},\mu,\nu;k,q\big) \\
	& \qquad \qquad - g_{2,I} \big((X_m)_{t+q}^{t+q+k-1},\mu,\nu;k,q\big) g_{3,R} \big((X_m)_{t}^{t+q+k-1},\mu,\nu;k,q\big) \\ 
	& \qquad \qquad  - g_{2,R} \big((X_m)_{t+q}^{t+q+k-1},\mu,\nu;k,q\big) g_{3,I} \big((X_m)_{t}^{t+q+k-1},\mu,\nu;k,q\big) \Big]^2 \bigg\}.
\end{align*}

Let $g_{Rm}$ and $g_{Im}$ denote
\begin{align*}
	g_{Rm} \Big( ( X_{j,m} )_{t}^{t+q+k-1}, \mu, \nu; k, q \Big) = & {g}_{1,R} \Big( ( X_{j,m} )_{t}^{t+q+k-1}, \mu, \nu; k, q \Big) \\
	& - {g}_{2,R} \Big(  X_{j,t+q+k-1}, \mu; k, q\Big) {g}_{3,R} \Big( ( X_{j,m} )_{t}^{t+q+k-1}, \nu; k, q \Big) \\
	& + {g}_{2,I} \Big(  X_{j,t+q+k-1}, \mu; k,q \Big) {g}_{3,I} \Big( ( X_{j,m} )_{t}^{t+q+k-1}, \nu; k,q \Big),
\end{align*}
and
\begin{align*}
	g_{Im} \Big( ( X_{j,m} )_{t}^{t+q+k-1}, \mu, \nu; k, q \Big) = & {g}_{1,I} \Big( ( X_{j,m} )_{t}^{t+q+k-1}, \mu, \nu; k, q \Big) \\
	& - {g}_{2,I} \Big(  X_{j,t+q+k-1}, \mu; k, q\Big) {g}_{3,R} \Big( ( X_{j,m} )_{t}^{t+q+k-1}, \nu; k, q \Big) \\
	& - {g}_{2,R} \Big(  X_{j,t+q+k-1}, \mu; k,q \Big) {g}_{3,I} \Big( ( X_{j,m} )_{t}^{t+q+k-1}, \nu; k,q \Big),
\end{align*}
respectively. And let $\hat{g}_{Rm}^{(-\mathcal{L})}$ and $\hat{g}_{Im}^{(-\mathcal{L})}$ denote the corresponding ``plug-in'' estimators with $\hat{g}_{r,R}^{(-\mathcal{L})}$ and $\hat{g}_{r,I}^{(-\mathcal{L})}$ ($r=1,2,3$).

Now that we can rewrite $\Gamma_{N,T}^{(k,q)}(\mu,\nu)$ as
\begin{align*}
    & \Gamma^{(k,q)}_{N,T}(\mu,\nu) \\
    = & \frac{1}{N_\mathcal{L}(T-q-k+1)}\sum_{j \in \mathcal{L}} \sum_{t=1}^{T-q-k+1} \bigg[ \hat{g}^{(-\mathcal{L})^2}_{Rm} \Big( ( X_{j,m} )_{t}^{t+q+k-1}, \mu, \nu; k, q \Big) \\
    & \qquad \qquad \qquad \qquad  \qquad \qquad \qquad \quad  + \hat{g}^{(-\mathcal{L})^2}_{Im} \Big( ( X_{j,m} )_{t}^{t+q+k-1}, \mu, \nu; k, q \Big) \bigg].
\end{align*}
We also define
\begin{align*}
    & \tilde{\Gamma}^{(k,q)}_{N,T}(\mu,\nu) \\
    = & \frac{1}{N_\mathcal{L}(T-q-k+1)}\sum_{j \in \mathcal{L}} \sum_{t=1}^{T-q-k+1} \bigg[ g^2_{Rm} \Big( ( X_{j,m} )_{t}^{t+q+k-1}, \mu, \nu; k, q \Big) \\
    & \qquad \qquad \qquad \qquad  \qquad \qquad \qquad \quad  + g^2_{Im} \Big( ( X_{j,m} )_{t}^{t+q+k-1}, \mu, \nu; k, q \Big) \bigg],
\end{align*}
with oracle functions $g_{Rm}$ and $g_{Im}$. Let
\begin{align*}
    & R_{N,T}^{(k,q)}(\mu,\nu) \\
    = & \Gamma_{N,T}{(k,q)}(\mu,\nu)-\tilde{\Gamma}_{N,T}{(k,q)}(\mu,\nu) \\
    = & \frac{1}{N_\mathcal{L}(T-q-k+1)}\\
    & \times \sum_{j \in \mathcal{L}} \sum_{t=1}^{T-q-k+1} \bigg[ \hat{g}^2_{Rm} \Big( ( X_{j,m} )_{t}^{t+q+k-1}, \mu, \nu; k, q \Big) + \hat{g}^2_{Im} \Big( ( X_{j,m} )_{t}^{t+q+k-1}, \mu, \nu; k, q \Big) \\
    & \qquad \qquad \qquad \quad  - g^2_{Rm} \Big( ( X_{j,m} )_{t}^{t+q+k-1}, \mu, \nu; k, q \Big) - g^2_{Im} \Big( ( X_{j,m} )_{t}^{t+q+k-1}, \mu, \nu; k, q \Big) \bigg],
\end{align*}
then
\begin{align*}
	& \max_{0 \le q \le Q} \max_{1 \le b \le B} \sqrt{N_\mathcal{L}(T-q-k+1)} \Gamma^{(k,q)}_{N,T}(\mu_b,\nu_b)\\
	= & \max_{0 \le q \le Q} \max_{1 \le b \le B} \sqrt{N_\mathcal{L}(T-q-k+1)} \Big[ \tilde{\Gamma}^{(k,q)}_{N,T}(\mu_b,\nu_b) + R^{(k,q)}_{N,T}(\mu_b,\nu_b) \Big].
\end{align*}

We can further decompose $R^{(k,q)}_{N,T}(\mu,\nu)$ into
\begin{equation*}
	 R^{(k,q)}_{N,T}(\mu,\nu) = R^{(k,q)}_{N,T,1}(\mu,\nu) + R^{(k,q)}_{N,T,2}(\mu,\nu) + R^{(k,q)}_{N,T,3}(\mu,\nu) + R^{(k,q)}_{N,T,4}(\mu,\nu) 
\end{equation*}
where
\begin{align*}
    & R^{(k,q)}_{N,T,1}(\mu,\nu) = \frac{1}{N_\mathcal{L}(T-q-k+1)}\sum_{j \in \mathcal{L}} \sum_{t=1}^{T-q-k+1} \bigg[ \hat{g}_{Rm} \Big( ( X_{j,m} )_{t}^{t+q+k-1}, \mu, \nu; k, q \Big) \\
    & \qquad \qquad \qquad \qquad \qquad \qquad  \qquad \qquad \qquad \qquad \quad - {g}_{Rm} \Big( ( X_{j,m} )_{t}^{t+q+k-1}, \mu, \nu; k, q \Big) \bigg]^2, \\
    & R^{(k,q)}_{N,T,2}(\mu,\nu) = \frac{1}{N_\mathcal{L}(T-q-k+1)}\sum_{j \in \mathcal{L}} \sum_{t=1}^{T-q-k+1} \bigg[ \hat{g}_{Im} \Big( ( X_{j,m} )_{t}^{t+q+k-1}, \mu, \nu; k, q \Big)\\
    & \qquad \qquad \qquad \qquad \qquad \qquad  \qquad \qquad \qquad \qquad \quad - {g}_{Im} \Big( ( X_{j,m} )_{t}^{t+q+k-1}, \mu, \nu; k, q \Big) \bigg]^2, \\
    & R^{(k,q)}_{N,T,3}(\mu,\nu) = \frac{1}{N_\mathcal{L}(T-q-k+1)}\sum_{j \in \mathcal{L}} \sum_{t=1}^{T-q-k+1} \Bigg\{ 2 {g}_{Rm}  \Big( ( X_{j,m} )_{t}^{t+q+k-1}, \mu, \nu; k, q \Big)\\
    & \qquad \qquad \qquad \qquad \times\bigg[ \hat{g}_{Rm} \Big( ( X_{j,m} )_{t}^{t+q+k-1}, \mu, \nu; k, q \Big) -  {g}_{Rm} \Big( ( X_{j,m} )_{t}^{t+q+k-1}, \mu, \nu; k, q \Big)\bigg]\Bigg\}, \\
    & R^{(k,q)}_{N,T,4}(\mu,\nu) = \frac{1}{N_\mathcal{L}(T-q-k+1)}\sum_{j \in \mathcal{L}} \sum_{t=1}^{T-q-k+1} \Bigg\{ 2 {g}_{Im}  \Big( ( X_{j,m} )_{t}^{t+q+k-1}, \mu, \nu; k, q \Big)\\
    & \qquad \qquad \qquad \qquad \times\bigg[ \hat{g}_{Im} \Big( ( X_{j,m} )_{t}^{t+q+k-1}, \mu, \nu; k, q \Big) -  {g}_{Im} \Big( ( X_{j,m} )_{t}^{t+q+k-1}, \mu, \nu; k, q \Big)\bigg]\Bigg\}.
\end{align*}

Remind that Equation (\ref{Equi_Appx}) holds when $k \geq k_0$. Therefore, 
\begin{equation*}
    {g}_{Rm}  \Big( ( X_{j,m} )_{t}^{t+q+k-1}, \mu, \nu; k, q \Big) = 0, \qquad {g}_{Im}  \Big( ( X_{j,m} )_{t}^{t+q+k-1}, \mu, \nu; k, q \Big) = 0,
\end{equation*}
for any $j$, $t$, $\mu$, $\nu$, and $q$. Moreover,
\begin{equation*}
	R^{(k,q)}_{N,T,3}(\mu,\nu) = 0, \qquad R^{(k,q)}_{N,T,4}(\mu,\nu) = 0,
\end{equation*}
for any $\mu$, $\nu$, and $q$. In other words, for $k \geq k_0$, to prove
\begin{equation}\label{remain}
	\max_{0 \le q \le Q} \max_{1 \le b \le B} \sqrt{N_\mathcal{L}(T-q-k+1)} \left\vert  R^{(k,q)}_{N,T}(\mu_b,\nu_b) \right\vert = o_p(1),
\end{equation}
it suffices to prove that
\begin{equation}\label{remain_sub}
	\max_{0 \le q \le Q} \max_{1 \le b \le B} \sqrt{N_\mathcal{L}(T-q-k+1)} \left\vert  R^{(k,q)}_{N,T,r}(\mu_b,\nu_b) \right\vert = o_p(1),
\end{equation}
for $r=1,2$. 

We first prove (\ref{remain_sub}) for $m=1$. The idea follows \cite{shi2020does}, which utilize Theorem 4.2 of \cite{chen2015optimal}. First, we can demonstrate that 
\begin{equation*}
    \hat{g}_{Rm} \Big( ( X_{j,m} )_{t}^{t+q+k-1}, \mu, \nu; k, q \Big) - {g}_{Rm} \Big( ( X_{j,m} )_{t}^{t+q+k-1}, \mu, \nu; k, q \Big)
\end{equation*}
are values of a function of a 1-order Markov process, respectively, for $j \in \mathcal{L}$, $1 \le t \le T-q-k+1$. For sequence $\{ X_t \}$, we can first prove that the sequence is a $k_0$-order markov process. Then we can reconstruct a 1-order Markov process in the following way. Let $\tilde{k} = \max \{ k_0, q+k \}$, and define new random variables
\begin{align*}
    X_{j,t}^{(\tilde{k})} = 
    \begin{cases}
        \left(X_{l_j,t}^\top, \cdots, X_{l_j,t+\tilde{k}-1}^\top \right)^\top, & 1 \le t \le T-\tilde{k}+1,\\
        \left(X_{l_j,t}^\top, \cdots, X_{l_j,T}^\top, X_{l_{j+1},1}^\top, \cdots,  X_{l_{j+1},\tilde{k}+t-T-1}^\top \right)^\top, & T-\tilde{k}+2 \le t \le T-q-k+1, 
    \end{cases}
\end{align*}
where $\mathcal{L}=\{ l_1,\cdots,l_{N_\mathcal{L}} \}$, and $X_{l_{n_{\mathcal{L}}+1},t} = \bm{0}$, for $t \geq 1$. In this way, we have defined a $\tilde{k}(p+1)$-dimensional random process: $\{ Y(j), 1 \le j \le N_\mathcal{L}(T-q-k+1) \}$, and
\begin{equation*}
    Y\Big(u(T-q-k+1)+v\Big) = X_{u,v}^{(\tilde{k})}, \qquad 1 \le u \le N_\mathcal{L}, \quad 1 \le v \le T-q-k+1.
\end{equation*}
Now $\left\{ \left\{ X_{j,t}^{(\tilde{k})}\right\}_{1 \leq t \leq T-q-k+1} \right\}_{1 \leq j \leq N_\mathcal{L}}$ are $N_\mathcal{L}$ independent realizations of a 1-order Markov sequence $\left\{ X_{0,t}^{(\tilde{k})} \right\}_{1 \leq t \leq T-q-k+1}$. Following from Theorem 3.7 of \cite{bradley2005basic}, under stationary and ergodicity assumptions, the reorganized sequence $\{ Y(j) \}_{1\le j \le N_\mathcal{L}(T-q-k+1)}$ is a $\beta$-mixing sequence with the mixing coefficient $\beta(t) = O\left( \rho^t \right)$, for some positive $\rho < 1$ and positive integer $t$. 

For a $\tilde{k}(p+1)$-dimensional random variable $\tilde{X}$, let $\tilde{X}(1)$ denote the vector consisting of the first $(q+k)(p+1)$ entries of $\tilde{X}$. Define a function as follows.
\begin{equation*}
	\Lambda_1 \Big(\tilde{X},\mu,\nu;k,q \Big)= \hat{g}_{Rm} \Big( \tilde{X}(1), \mu, \nu; k, q \Big) - {g}_{Rm} \Big( \tilde{X}(1), \mu, \nu; k, q \Big).
\end{equation*}
Now for any fixed $1 \leq j \leq N_\mathcal{L}$, $1 \leq t \leq T-q-k+1$,
\begin{equation*}
	\hat{g}_{Rm} \Big( ( X_{l_j,m} )_{t}^{t+q+k-1}, \mu, \nu; k, q \Big) - {g}_{Rm} \Big( ( X_{l_j,m} )_{t}^{t+q+k-1}, \mu, \nu; k, q \Big) = \Lambda_1 \Big( X_{j,t}^{(\tilde{k})}, \mu, \nu; k,q \Big).
\end{equation*}
Now we have proven that $\hat{g}_{Rm} \Big( ( X_{j,m} )_{t}^{t+q+k-1}, \mu, \nu; k, q \Big) - {g}_{Rm} \Big( ( X_{j,m} )_{t}^{t+q+k-1}, \mu, \nu; k, q \Big)$ are values of a function of $\left\{ \left\{X_{j,t}^{(\tilde{k})}\right\}_{1 \leq t \leq T-q-k+1} \right\}_{1 \leq j \leq N_\mathcal{L}}$ or $\{ Y(j) \}_{1\leq j \leq N_\mathcal{L}(T-q-k+1)}$.

To employ the Theorem 4.2 of \cite{chen2015optimal}, we then provide some boundness properties of $\Lambda_1$. Let
\begin{equation*}
	\Phi^{(k,q)}_{j,t,b,1} = \Lambda_1\Big( (X_{l_j,m} )_{t}^{t+q+k-1}, \mu_b, \nu_b;k,q \Big) =  \Lambda_1\Big( Y\big(j(T-q-k+1)+t\big), \mu_b, \nu_b;k,q \Big).
\end{equation*}
By the stationary condition, we can derive that
\begin{equation}
	\max_{j,t,b} \mathrm{E}_{Y\left( j(T-q-k+1)+t \right)}\left[ \Phi^{(k,q)}_{j,t,b,1} \right]^4\leq 36 \max_{1 \leq b \leq B} \int_x \bigg[ \hat{g}_{Rm} \Big( x, \mu, \nu; k, q \Big) - {g}_{Rm}  \Big( x, \mu, \nu; k, q \Big) \bigg]^2 dF(x), \label{4moment1}
\end{equation}
where the expectation $\mathrm{E}_{Y\left( j(T-q-k+1)+t \right)}$ is taken with respect to $Y\left( j(T-q-k+1)+t \right)$. Let 
\begin{equation*}
    \Delta_1 := 36 \max_{1 \leq b \leq B} \int_x \bigg[ \hat{g}_{Rm} \Big( x, \mu, \nu; k, q \Big) - {g}_{Rm}  \Big( x, \mu, \nu; k, q \Big) \bigg]^2 dF(x)
\end{equation*}
It is worthwhile to mention that $\Delta_1$ is a random variable that depends on $\{ \mu_b, \nu_b\}_{b=1}^B$ and $\{ X_{j,t} \}_{j \notin \mathcal{L}, 1 \leq t \leq T}$. By Inequality (\ref{4moment1}), we can derive that
\begin{equation*}
	\max_{j,t,b} \mathrm{E}_{Y\left( j(T-q-k+1)+t \right)} \left[ \left( \Phi^{(k,q)}_{j,t,b,1} \right)^2 - \mathrm{E}_{Y\left( j(T-q-k+1)+t \right)} \left( \Phi^{(k,q)}_{j,t,b} \right)^2 \right]^2 \leq \Delta_1.
\end{equation*}
 Notice that the oracle and estimated conditional characteristic functions are all bounded, as $\left\vert \Phi^{(k,q)}_{j,t,b,1} \right\vert \leq 6$ and hence $\left\vert \left( \Phi^{(k,q)}_{j,t,b,1} \right)^2 - \mathrm{E}_{Y\left( j(T-q-k+1)+t \right)} \left( \Phi^{(k,q)}_{j,t,b,1} \right)^2 \right\vert \leq 36$. 

 According to Theorem 4.2 of \cite{chen2015optimal}, for any integers $\tau \geq 0$ and $1 < d < N_\mathcal{L}T/2$,
\begin{align*}
    & \mathrm{Pr}\left( \left\vert \sum_{j \in \mathcal{L}} \sum_{t=1}^{T-q-k+1}  \left( \Phi^{(k,q)}_{j,t,b,1} \right)^2 - \mathrm{E}_{Y\left( 0 \right)} \left( \Phi^{(k,q)}_{j,t,b,1} \right)^2 \right\vert \geq 6\tau \, \Bigg\vert \, \Delta_1 \right) \\
    \leq & \frac{N_\mathcal{L}(T-q-k+1)}{d} \beta(d) + \mathrm{Pr} \left( \left\vert \sum_{(j,t) \in \mathcal{L}_r} \left( \Phi^{(k,q)}_{j,t,b,1} \right)^2 - \mathrm{E}_{Y\left( 0 \right)} \left( \Phi^{(k,q)}_{j,t,b,1} \right)^2 \right\vert \geq \tau \, \Bigg\vert \, \Delta_1  \right) \\
	& + 4 \exp\left( - \frac{\tau^2/2}{N_\mathcal{L}(T-q-k+1)d\Delta_1 + 6d\tau/3} \right),
\end{align*}
where $\mathcal{L}_r$ denotes the last $N_\mathcal{L}(T-q-k+1) - d\lfloor N_\mathcal{L}(T-q-k+1)/d \rfloor$ elements in the list
\begin{equation*}
	\{ (l_1,1),\cdots,(l_1,T-q-k+1),(l_2,1),\cdots, (l_{N_\mathcal{L}},1), \cdots, (l_{N_\mathcal{L}},T-q-k+1) \}.
\end{equation*}
Suppose $\tau \geq 36d$, it is obvious that
\begin{equation*}
	\mathrm{Pr} \left( \left\vert \sum_{(j,t) \in \mathcal{L}_r} \left( \Phi^{(k,q)}_{j,t,b,1} \right)^2 - \mathrm{E}_{Y\left( 0 \right)} \left( \Phi^{(k,q)}_{j,t,b,1} \right)^2 \right\vert \geq \tau \,\Bigg\vert \, \Delta_1  \right) = 0.
\end{equation*}
Notice that $\beta(t) = O(\rho^t)$. Set $d = -(c^*+4)\ln(N_\mathcal{L}T)/\ln \rho$, we can derive that
\begin{equation*}
	N_\mathcal{L}(T-q-k+1) \beta(d)/d = O\left( K^{-1}Q^{-1}B^{-1}N_\mathcal{L}^{-1}T^{-1} \right),
\end{equation*}
for $Q+K \leq T$ and $B = O \left( (N_\mathcal{L}T)^{c^\star} \right)$. Notice that this asymptotic boundness property is uniform for any $1 \leq b \leq B$, $0 \leq q \leq Q$, and $1 \leq k \leq K$. Let $\tau$ takes value as
\begin{equation*}
	\tau = \max \left\{ 2 \sqrt{N_\mathcal{L}T d \Delta_1[\ln B + 3 \ln(N_\mathcal{L}T)]}, 8d[\ln B + 3 \ln(N_\mathcal{L}T)] \right\},
\end{equation*}
for sufficiently large $N_\mathcal{L}$ and $T$, it can be guaranteed that $\tau > 36d$. Hence, we can derive that
\begin{equation*}
	4 \exp\left( - \frac{\tau^2/2}{N_\mathcal{L}(T-q-k+1)d\Delta_1 + 6d\tau/3} \right) = O\left( K^{-1}Q^{-1}B^{-1}N_\mathcal{L}^{-1}T^{-1} \right).
\end{equation*}
Altogether, we have obtained that
\begin{equation*}
	\mathrm{Pr}\left( \left\vert \sum_{j \in \mathcal{L}} \sum_{t=1}^{T-q-k+1}  \left( \Phi^{(k,q)}_{j,t,b,1} \right)^2 - \mathrm{E}_{Y\left( 0 \right)} \left( \Phi^{(k,q)}_{j,t,b,1} \right)^2 \right\vert  \geq 6\tau \, \Bigg\vert \, \Delta_1 \right) = O\left( K^{-1}Q^{-1}B^{-1}N_\mathcal{L}^{-1}T^{-1} \right),
\end{equation*}
and
\begin{equation*}
	\mathrm{Pr}\left( \max_{k_0 \leq k\leq K} \max_{0 \leq q \leq Q} \max_{1 \leq b \leq B}  \left\vert \sum_{j \in \mathcal{L}} \sum_{t=1}^{T-q-k+1}  \left( \Phi^{(k,q)}_{j,t,b,1} \right)^2 - \mathrm{E}_{Y\left( 0 \right)} \left( \Phi^{(k,q)}_{j,t,b,1} \right)^2 \right\vert  \geq 6\tau \, \Bigg\vert \, \Delta_1 \right)  = O\left( N_\mathcal{L}^{-1}T^{-1} \right),
\end{equation*}
under the given condition on $Q$ and $K$, $T-q-k+1$ stays proportional to $T$. We can derive that, with probability $1-O\left( N_\mathcal{L}^{-1}T^{-1} \right)$,
\begin{equation*}
	\max_{0 \le q \le Q} \max_{1 \le b \le B} \sqrt{N_\mathcal{L}(T-q-k+1)} \left\vert  R^{(k,q)}_{N,T,1}(\mu_b,\nu_b) \right\vert = O\left( \sqrt{\Delta_1 \ln(N_\mathcal{L}T)}, \ln(N_\mathcal{L}T)/\sqrt{N_\mathcal{L}T}\right),
\end{equation*}
for any $k_0 \leq k \leq K$. Together with the convergence rate assumption of $\Delta_1$, we can derive that (\ref{remain_sub}) holds for $m=1$. For $m=2$, (\ref{remain_sub}) can be proved following the similar idea. Altogether, we have derived that, for $k \geq k_0$,
\begin{equation*}
	\max_{0 \le q \le Q} \max_{1 \le b \le B} \sqrt{N_\mathcal{L}(T-q-k+1)} \left\vert  R^{(k,q)}_{N,T}(\mu_b,\nu_b) \right\vert = o_p(1).
\end{equation*}
Therefore,
\begin{align*}
	& \max_{0 \le q \le Q} \max_{1 \le b \le B} \sqrt{N_\mathcal{L}(T-q-k+1)} {\Gamma}^{(k,q)}_{N,T}(\mu_b,\nu_b)\\
	 = & \max_{0 \le q \le Q} \max_{1 \le b \le B} \sqrt{N_\mathcal{L}(T-q-k+1)} \tilde{\Gamma}^{(k,q)}_{N,T}(\mu_b,\nu_b) + o_p(1).
\end{align*}
Moreover, for $k \geq k_0$, $\tilde{\Gamma}^{(k,q)}_{N,T}(\mu_b,\nu_b) = 0$ for any $q$ and $(\mu_b,\nu_b)$. Then, \textcolor{blue}{check the presentation, whether we should use the 'converge in probability'}
\begin{equation*}
	\max_{0 \le q \le Q} \max_{1 \le b \le B} \sqrt{N_\mathcal{L}(T-q-k+1)} {\Gamma}^{(k,q)}_{N,T}(\mu_b,\nu_b) \stackrel{d}{\longrightarrow} 0.
\end{equation*}

For the next part, we will derive the convergence rate of $\Pi^{(k)}_{N,T}$ converging to $R^{(k)}$ for any $k$. Remind that
\begin{equation*}
	\Pi^{(k)}_{N,T} = \max_{0 \le q \le Q} \max_{1 \le b \le B}\Gamma^{(k,q)}_{N,T}(\mu_b,\nu_b),
\end{equation*}
and
\begin{equation*}
	R^{(k)} = \max_{0 \leq q \leq Q} \sup_{\mu \in \mathbb{R}^p, \nu \in \mathbb{R}^{p+1}} \mathrm{E}\left[ \Gamma^{(k,q)}(\mu,\nu) \right].
\end{equation*}
We can also rewrite $\Pi_{N,T}^{(k)}$ as
\begin{equation*}
	\Pi_{N,T}^{(k)} = \max_{0 \leq q \leq Q} \max_{1 \leq b \leq B} \left[ \tilde{\Gamma}_{N,T}^{(k,q)}(\mu_b,\nu_b) + R_{N,T}^{(k,q)}(\mu_b,\nu_b) \right].
\end{equation*}
Following the same method aforementioned, we can prove that
\begin{equation*}
	\max_{0 \leq q \leq Q} \max_{1 \leq b \leq B} \left\vert R_{N,T,r}^{(k,q)}(\mu_b,\nu_b) \right\vert = o_p\left( \sqrt{\frac{\ln(N_\mathcal{L}T)}{N_\mathcal{L}T} }\right),
\end{equation*}
for $r=1,2,3,4$.

For $\max_{0 \leq q \leq Q} \max_{1 \leq b \leq B}  \tilde{\Gamma}_{N,T}^{(k,q)}(\mu_b,\nu_b)$, we try to utilize Theorem 4.2 of \cite{chen2015optimal}. Remind $\left\{ X_{j,t}^{(\tilde{k})} \right\}_{1 \leq j \leq N_\mathcal{L}}$ and $\{ Y(j) \}_{1\le j \le N_\mathcal{L}(T-q-k+1)}$ aforementioned. For a $\tilde{k}(p+1)$-dimensional random variable $\tilde{X}$, let $\tilde{X}(1)$ denote the vector consisting of the first $(q+k)(p+1)$ entries of $\tilde{X}$,  and $\tilde{X}(2)$ denote the vector consisting of the $[(q+k-1)(p+1)+1]$-th to the $(q+k)(p+1)$-th entries. We can define a function as follows.
\begin{equation*}
	\Psi \big( \tilde{X},\mu,\nu;k,q \big) := g^2_{Rm} \left(\tilde{X}(1),\mu,\nu;k,q \right) + g^2_{Im} \left(\tilde{X}(1),\mu,\nu;k,q \right).
\end{equation*}
Let $\Xi \big( \tilde{X},\mu,\nu;k,q \big) = \Psi \big( \tilde{X},\mu,\nu;k,q\big) - \mathrm{E}_{\tilde{X}} [ \Psi \big( \tilde{X},\mu,\nu;k,q \big)] $. Given $k$, $q$, $\mu_b$ and $\nu_b$, we can further derive a random variable array as $\Xi_{j,t}(\mu_b,\nu_b;k,q) := \Xi \big( X_{j,t}^{(k)}, \mu_b, \nu_b; k, q \big)$. Then
\begin{equation*}
	N_{\mathcal{L}} (T-q-k+1) \Big[ \tilde{\Gamma}_{N,T}^{(k,q)} (\mu_b,\nu_b) - \mathrm{E} \big[ \tilde{\Gamma}_{N,T}^{(k,q)} (\mu_b,\nu_b)\big] \Big] = \sum_{j=1}^{N_\mathcal{L}} \sum_{t=1}^{T-q-k+1}\Xi_{j,t}(\mu_b,\nu_b;k,q)
\end{equation*}
It is easy to verify that
\begin{equation*}
	\max_{j,t} \vert \Xi_{j,t}(\mu_b,\nu_b;k,q) \vert \le 18,
\end{equation*}
and
\begin{equation*}
	\max_{j_1,j_2,t_1,t_2} \mathbb{E} \left[ \Xi_{j_1,t_1}(\mu_b,\nu_b;k,q) \Xi_{j_2,t_2}(\mu_b,\nu_b;k,q) \right] \le \max_{j,t} \mathbb{E} \left[ \Xi^2_{j,t}(\mu_b,\nu_b;k,q) \right] \le 324.
\end{equation*}

With the mixing property of sequence $\{ Y(j) \}$ and the properties of $\Xi\big( \cdot, \mu_b, \nu_b;k,q \big)$, we can further utilize Theorem 4.2 of \cite{chen2015optimal}. Given $q$, $k$, $\{\mu_b\}_{b=1}^B$ and $\{\nu_b\}_{b=1}^B$, we can bound the tail property as
\begin{align*}
	& \mathrm{Pr} \left( \bigg\vert \sum_{j=1}^{N_\mathcal{L}} \sum_{t=1}^{T-q-k+1} \Xi_{j,t}\big( \mu_b, \nu_b;k,q \big) \bigg\vert \ge 6 \tau \, \Bigg\vert \, \{ \mu_b \}_{b=1}^B, \{ \nu_b \}_{b=1}^B \right)
	\\
	\le & \frac{N_{\mathcal{L}} (T-q-k+1)}{d} \beta(d) + \mathrm{Pr} \left( \bigg\vert \sum_{(j,t) \in \mathcal{I}_d} \Xi_{i,t}\big( \mu_b, \nu_b;k,q \big) \bigg\vert \ge \tau \, \Bigg\vert \,  \{ \mu_b \}_{b=1}^B, \{ \nu_b \}_{b=1}^B   \right)
	\\
	& + 4 \exp \left( -\frac{\tau^2/2}{324 N_\mathcal{L}(T-q-k+1) d + 18 d\tau/3} \right),
\end{align*}
where $\mathcal{I}_d$ consists of the last $N_\mathcal{L} (T-q-k+1) - \lfloor N_\mathcal{L} (T-q-k+1)/d \rfloor d$ elements (for some $d > 0$) of the following list:
\begin{equation*}
	(1,1), \cdots, (1, T-q-k+1), \cdots, (N_\mathcal{L},1), \cdots, (N_\mathcal{L},T-q-k+1).
\end{equation*}
Set $d=-(c^*+4) \ln(N_\mathcal{L}T)/\ln \rho$. We can derive that
\begin{equation*}
	\frac{N_\mathcal{L}(T-q-k+1)}{d} \beta(d) = O(B^{-1} Q^{-1} K^{-1} N_\mathcal{L}^{-1} T^{-1}).
\end{equation*}
Let $\tau = \max \{ 36 \sqrt{3 N_\mathcal{L}Td\ln(BN_\mathcal{L}T)}, 72 d\ln(BN_\mathcal{L}T) \}$. We can obtain that
\begin{equation*}
	4 \exp \left( -\frac{\tau^2/2}{324 N_\mathcal{L}(T-q-k+1) d + 18 d\tau/3} \right) \le 4 (BN_\mathcal{L}T)^{-3} = O(B^{-1} Q^{-1} K^{-1} N_\mathcal{L}^{-1} T^{-1}) .
\end{equation*}
Note that $N_\mathcal{L}(T-q-k+1) - \lfloor N_\mathcal{L}(T-q-k+1)/d  \rfloor d \le d$ and $\tau > 18d$ for sufficiently large $N_\mathcal{L}$ or $T$. It is obvious that
\begin{equation*}
	\mathrm{Pr} \left( \bigg\vert \sum_{(j,t) \in \mathcal{I}_d} \Xi_{j,t}\big( \mu_b, \nu_b;k,q \big) \bigg\vert \ge \tau \, \Bigg\vert \,  \{ \mu_b \}_{b=1}^B, \{ \nu_b \}_{b=1}^B \right) = 0.
\end{equation*}
Now we can derive that
\begin{align*}
	& \mathrm{Pr} \left(N_\mathcal{L}(T-q-k+1) \left| \Gamma_{N,T}^{(k,q)} (\mu_b,\nu_b) - \mathrm{E}\big[ \Gamma_{N,T}^{(k,q)} (\mu_b,\nu_b) \big] \right| \ge 6\tau \, \bigg\vert \, \{ \mu_b \}_{b=1}^B, \{ \nu_b \}_{b=1}^B  \right) \\
	= & \, \mathrm{Pr} \left( \left\vert \sum_{j=1}^{N_\mathcal{L}} \sum_{t=1}^{T-q-k+1}\Xi_{j,t}(\mu_b,\nu_b;k,q) \right\vert \ge 6\tau  \, \bigg\vert \, \{ \mu_b \}_{b=1}^B, \{ \nu_b \}_{b=1}^B   \right) = O \left( B^{-1} Q^{-1} K^{-1} N_\mathcal{L}^{-1} T^{-1} \right),
\end{align*}
uniformly for any $0 \le q \le Q$, $1 \le b \le B$ and $1 \le k \le K$.  
 
Then there exists $\kappa>0$ such that
\begin{align*}
	& \mathrm{Pr} \left( \max_{1 \le k \le K} \left\vert \Pi^{(k)}_{N,T} - R^{(k)} \right\vert \ge \kappa \right) \\
	= & \mathrm{Pr} \Bigg( \max_{1 \le k \le K} \bigg\vert \Pi^{(k)}_{N,T} - \max_{0 \leq q \leq Q} \max_{1 \leq b \leq B} \mathrm{E}\left[ \Gamma^{(k,q)}(\mu_b,\nu_b) \right] \\
    & \qquad \qquad \qquad + \max_{0 \leq q \leq Q} \max_{1 \leq b \leq B} \mathrm{E}\left[ \Gamma^{(k,q)}(\mu_b,\nu_b) \right] - R^{(k)} \bigg\vert \ge \kappa \Bigg) \\
	\leq &  \mathrm{Pr} \left( \max_{1 \le k \le K} \left\vert \Pi^{(k)}_{N,T} - \max_{0 \leq q \leq Q} \max_{1 \leq b \leq B} \mathrm{E}\left[ \Gamma^{(k,q)}(\mu_b,\nu_b) \right] \right\vert \ge \kappa/2 \right) \\
	& +  \mathrm{Pr} \left( \max_{1 \le k \le K} \left\vert \max_{0 \leq q \leq Q} \max_{1 \leq b \leq B} \mathrm{E}\left[ \Gamma^{(k,q)}(\mu_b,\nu_b) \right] - R^{(k)} \right\vert \ge \kappa/2 \right).
\end{align*}
Here
\begin{equation*}
	 \mathrm{Pr} \left( \max_{1 \le k \le K} \left\vert \max_{0 \leq q \leq Q} \max_{1 \leq b \leq B} \mathrm{E}\left[ \Gamma^{(k,q)}(\mu_b,\nu_b) \right] - R^{(k)} \right\vert \ge \kappa/2 \right) = O(N_\mathcal{L}^{-1}T^{-1}),
\end{equation*}
and
\begin{align*}
	& \mathrm{Pr} \left( \max_{1 \le k \le K} \left\vert \Pi^{(k)}_{N,T} - R^{(k)} \right\vert \ge \kappa/2 \, \bigg\vert \, \{\mu_b\}_{b=1}^B, \{\nu_b\}_{b=1}^B \right) \\
	= & \mathrm{Pr} \left( \max_{1 \le k \le K} \left\vert \Pi^{(k)}_{N,T} - \max_{0 \le q \le Q} \max_{1 \le b \le B} \mathrm{E} \big[ \Gamma^{(k,q)}(\mu_b,\nu_b) \big] \right\vert \ge \kappa/2 \, \bigg\vert \, \{\mu_b\}_{b=1}^B, \{\nu_b\}_{b=1}^B  \right) \\
	\le & \mathrm{Pr} \bigg( \max_{1 \le k \le K} \max_{0\le q \le Q} \max_{1 \le b \le B}\Big\vert \tilde{\Gamma}^{(k,q)}_{N,T}(\mu_b,\nu_b) + R_{N,T}^{(k,q)}(\mu_b,\nu_b) \\
    & \qquad \qquad \qquad \qquad \qquad \quad - \mathrm{E} \big[ \Gamma^{(k,q)}_{N,T}(\mu_b,\nu_b) \big]  \Big\vert \ge \kappa/2  \, \bigg\vert \, \{\mu_b\}_{b=1}^B, \{\nu_b\}_{b=1}^B \bigg) 
	\\
	\le & \sum_{k=1}^K \sum_{q=0}^Q \sum_{b=1}^B \left[ \mathrm{Pr} \left(\left\vert \Gamma^{(k,q)}_{N,T}(\mu_b,\nu_b) - \mathrm{E} \big[ \Gamma^{(k,q)}_{N,T}(\mu_b,\nu_b) \big]  \right\vert \ge \kappa/4 \, \bigg\vert \, \{\mu_b\}_{b=1}^B, \{\nu_b\}_{b=1}^B \right) \right. \\
	& \qquad \qquad \qquad \left. + \mathrm{Pr} \left(\left\vert R^{(k,q)}_{N,T}(\mu_b,\nu_b) \right\vert \ge \kappa/4 \, \bigg\vert \, \{\mu_b\}_{b=1}^B, \{\nu_b\}_{b=1}^B \right) \right]
	\\
	= & O(N_\mathcal{L}^{-1}T^{-1}).
\end{align*}
Here $\kappa = \max\left\{ O\left(\sqrt{\frac{\ln(N_\mathcal{L}T)}{N_\mathcal{L}T}}\right), O\left(\frac{\ln(N_\mathcal{L}T)}{N_\mathcal{L}T}\right) \right\}$. So we can claim that 
\begin{equation*}
	\Big\vert \Pi^{(k)}_{N,T} - R^{(k)} \Big\vert = O_p\left( \sqrt{\frac{\ln(N_\mathcal{L}T)}{N_\mathcal{L}T}} \right) = O_p\left( \sqrt{\frac{\ln(NT)}{NT}} \right)
\end{equation*}
for any $1 \le k \le K$.


\subsection{Proof of Theorem \ref{consistency}}

In this subsection, we will derive the consistency of our estimator, that is,
\begin{equation*}
	\mathrm{Pr} \big(\hat{k}_\tau = k_0 \big) \to 1, \qquad N, T \to \infty.
\end{equation*}
Notice that 
\begin{equation*}
	\mathrm{Pr} \big(\hat{k}_\tau = k_0 \big) = 1 - \mathrm{Pr} \big( \hat{k}_\tau \le k_0 - 1 \big) - \mathrm{Pr} \big( \hat{k}_\tau \ge k_0 + 1 \big).
\end{equation*}
For the first subtrahend, notice that if $k_0 = 1$, $\mathrm{Pr} \big( \hat{k}_\tau \le k_0 - 1 \big)=0$. If $k_0 > 1$, then
\begin{align*}
	& \mathrm{Pr} \Big( \hat{k}_\tau \le k_0 - 1 \Big) = \mathrm{Pr} \Big( \tilde{\Omega}_{N,T}^{(k_0)} > \tau \Big) = \mathrm{Pr} \left( \frac{\left(\Pi_{N,T}^{(k_0)}\right)^\eta + \tilde{c}_{N,T}}{\left(\Pi_{N,T}^{(k_0-1)}\right)^\eta + \tilde{c}_{N,T}} > \tau \right) \\
	= & \mathrm{Pr} \Bigg\{ \left(\Pi_{N,T}^{(k_0)}\right)^\eta + \tilde{c}_{N,T} > \tau \left[ \left(\Pi_{N,T}^{(k_0-1)}\right)^\eta + \tilde{c}_{N,T} \right] \Bigg\} \\
	= & \mathrm{Pr} \left\{ O_p\left(  \left( \frac{\ln(N_\mathcal{L}T)}{N_\mathcal{L}T} \right)^{\eta/2} \right) + \tilde{c}_{N,T} > \tau \left[ \left(R^{(k_0-1)}\right)^\eta + O_p\left(  \left( \frac{\ln(N_\mathcal{L}T)}{N_\mathcal{L}T} \right)^{1/2} \right) + \tilde{c}_{N,T} \right] \right\} \\
	= & \mathrm{Pr} \left\{ \tau \left(R^{(k_0-1)}\right)^\eta <  O_p\left(  \left( \frac{\ln(N_\mathcal{L}T)}{N_\mathcal{L}T} \right)^{1/2} \right) + (1-\tau)  \tilde{c}_{N,T} \right\} \to 0.
\end{align*}
For the second subtrahend, notice that there exists at least one integer $\tilde{k}$ such that $k_0+1 \le \tilde{k} \le \hat{k}_\tau$, and
\begin{align*}
	& \mathrm{Pr} \Big( \hat{k}_\tau \ge k_0 + 1 \Big) \le \mathrm{Pr} \Big( \tilde{\Omega}_{N,T}^{(\tilde{k})} \le \tau \Big) = \mathrm{Pr} \left( \frac{\left(\Pi_{N,T}^{(\tilde{k})}\right)^\eta + \tilde{c}_{N,T}}{\left(\Pi_{N,T}^{(\tilde{k}-1)}\right)^\eta + \tilde{c}_{N,T}} \le \tau \right) \\
	= & \mathrm{Pr} \left\{ O_p\left(  \left( \frac{\ln(N_\mathcal{L}T)}{N_\mathcal{L}T} \right)^{\eta/2} \right) + \tilde{c}_{N,T} \le \tau \left[O_p\left(  \left( \frac{\ln(N_\mathcal{L}T)}{N_\mathcal{L}T} \right)^{\eta/2} \right) + \tilde{c}_{N,T} \right] \right\} \\
	= & \mathrm{Pr} \Bigg\{ 1 - \tau  \le O_p\left( \frac{[\ln(N_\mathcal{L}T)]^{\eta/2}}{\tilde{c}_{N,T} [N_\mathcal{L}T]^{\eta/2} }\right) \Bigg\} \to 0.
\end{align*}
Based on these two deductions, we can derive the consistency.

\section{Supplement of Section \ref{RealData}}

In Section \ref{RealData}, we have conducted a numerical studies based on a synthetic Markov decision process mimicking the OhioT1DM dataset. The model of the corresponding MDP is set as follows. Let $S_{1,t}$, $S_{2,t}$, and $S_{3,t}$ denote the average glucose level, average carbohydrate intake, and average exercise intensity at $t$. For any $t$, $S_{2,t}$ and $S_{3,t}$ are generated as follows.
\begin{equation*}
    I_{2,t} \sim B(1,0.1), \qquad I_{3,t} \sim B(1,0.015),
\end{equation*}
\begin{equation*}
    S_{2,t} \sim \begin{cases}
        \mathbf{1}_0, & I_{2,t}=0;\\
        \chi^2(10), & I_{2,t}=1;
    \end{cases}\qquad
    S_{3,t} \sim \begin{cases}
        \mathbf{1}_0, & I_{3,t}=0;\\
        Pois(5), & I_{3,t}=1,
    \end{cases}
\end{equation*}
where $\mathbf{1}_0$ denotes the point mass distribution at $0$.
$A_t$ is generated as
\begin{equation*}
    \mathrm{Pr}(A_t=j) = \begin{cases}
        0.8, & j=0;\\
        0.155, & j=1;\\
        0.03, & j=2;\\
        0.01, & j=3;\\
        0.005, & j=4.
    \end{cases}
\end{equation*}
Let $X_{t} = (S_{1,t},S_{2,t},S_{3,t},A_t)^\top$. Furthermore, $S_{1,t}$ is generated as
\begin{equation*}
    S_{1,t+2} = \begin{pmatrix}
        -0.377 \\ 0.165 \\ 0.329 \\ -5.271
    \end{pmatrix}^\top X_{t} + \begin{pmatrix}
        1.145 \\ 0.3 \\ -4.388 \\ -1.387
    \end{pmatrix}^\top X_{t+1} + \varepsilon_{S,1},
\end{equation*}
where $\varepsilon_{S,1}\sim\mathcal{N}(0,1)$.








\vskip 0.2in
\bibliography{ref_main}

\end{document}